\documentclass{article}
\usepackage{arxiv}
\usepackage[utf8]{inputenc} 
\usepackage[T1]{fontenc}    
\usepackage{hyperref}       
\usepackage{url}            
\usepackage{booktabs}       
\usepackage{amsfonts}       
\usepackage{nicefrac}       
\usepackage{microtype}      
\usepackage{lipsum}
\usepackage{xcolor}
\usepackage{amsmath,amstext,amssymb}
\usepackage{graphicx,verbatim,float,subfigure,graphicx}
\usepackage{threeparttable,amsmath,appendix}
\graphicspath{ {./images/} }

\usepackage{amssymb}
\usepackage{svg}
\usepackage{multirow}  
\usepackage[export]{adjustbox}
\usepackage{soul}
\usepackage{textcomp}
\usepackage{multirow}
\usepackage{enumitem}

\DeclareUnicodeCharacter{2061}{}

\title{Moisture Diffusion in Multi-Layered Materials: The Role of Layer Stacking and Composition}

\author{
  Shaojie Zhang \\
  Department of Civil \& Environmental Engineering \\
  University of Wisconsin-Madison, WI 53706 \\ 
\\
  Department of Civil Engineering\\
  Tsinghua University, Beijing, China
   \And
 Yuhao Liu\\
 Department of Mechanical Engineering \\
  University of Wisconsin-Madison, WI 53706 \\
  \And
Peng Feng \\
Department of Civil Engineering \\
Tsinghua University, Beijing, China
 \And
 Pavana Prabhakar \\
 Department of Mechanical Engineering \\
  Department of Civil \& Environmental Engineering \\
  University of Wisconsin-Madison, WI 53706 \\
  \texttt{pavana.prabhakar@wisc.edu} \\
}

\date{ }

\begin{document}
\maketitle
\begin{abstract}

Multi-layered materials are everywhere, from fiber-reinforced polymer composites (FRPCs) to plywood sheets to layered rocks. When in service, these materials are often exposed to long-term environmental factors, like moisture, temperature, salinity, etc. Moisture, in particular, is known to cause significant degradation of materials like polymers, often resulting in loss of material durability.
Hence, it is critical to determine the total diffusion coefficient of multi-layered materials given the coefficients of individual layers. However, the relationship between a multi-layered material's total diffusion coefficient and the individual layers' diffusion coefficients is not well established. Existing parallel and series models to determine the total diffusion coefficient do not account for the order of layer stacking. In this paper, we introduce three parameters influencing the diffusion behavior of multi-layered materials: the ratio of diffusion coefficients of individual layers, the volume fraction of individual layers, and the stacking order of individual layers. Computational models are developed within a finite element method framework to conduct parametric analysis considering the proposed parameters. We propose a new model to calculate the total diffusion coefficient of multi-layered materials more accurately than current models. We verify this parametric study by performing moisture immersion experiments on multi-layered materials. Finally, we propose a methodology for designing and optimizing the cross-section of multi-layered materials considering long-term moisture resistance. This study gives new insights into the diffusion behavior of multi-layered materials, focusing on polymer composites.

\end{abstract}


\section{Introduction}\label{intro}

Multi-layered materials, like fiber-reinforced polymer composites (FRPCs), are increasingly used in civil, marine, and aerospace engineering due to their high strength, lightweight, and corrosion-resistant properties\cite{VanDenEinde2003,Rubino2020,Nunes2016}. In service, FRPCs are exposed to harsh environmental factors such as water, dry and wet cycles, ultraviolet radiation, salt corrosion, acid and alkali solutions, high and low temperatures, and long-term loading\cite{Feng2016,Harries2017,Lu2015,Wu2021}. Environmental factors interact with each other and have irreversible effects on the long-term performance of FRPCs. Moisture ingression is a significant factor regardless of whether a composite is exposed to a humid atmosphere, acidic solution, or alkaline solution immersion\cite{Liu2020,PSubramaniyan2021}. Water molecules can enter the FRPC through diffusion, infiltration, and adsorption in water environments and interact with the FRPC material, degrading its properties.\cite{Birger1989,Feng2022,Wood1997}. 

Moisture transport in FRPCs is primarily dominated by diffusion. Studies on moisture diffusion in composite materials typically focus on experiments or simulations\cite{PSubramaniyan2021}. Both experimental and simulation studies are based on moisture diffusion kinetics, and the standard diffusion models used are the Fickian and Non-Fickian diffusion models \cite{Bond2005}. The Fick diffusion model is the most widely used model for characterizing moisture diffusion in composite materials, and it is based on Fick's second law. Studies have shown that several models have been proposed to describe composite materials' anomalous water uptake behavior, including the Langmuir model \cite{Carter1978} and the Two-stage model \cite{Bagley1955}. Most of these models are based on the definition and mechanism of the Fickian Diffusion Model. Therefore, this paper will still focus on the Fickian Model.

\begin{figure}[h!]  
     \centering
     \includegraphics[width=1\linewidth]{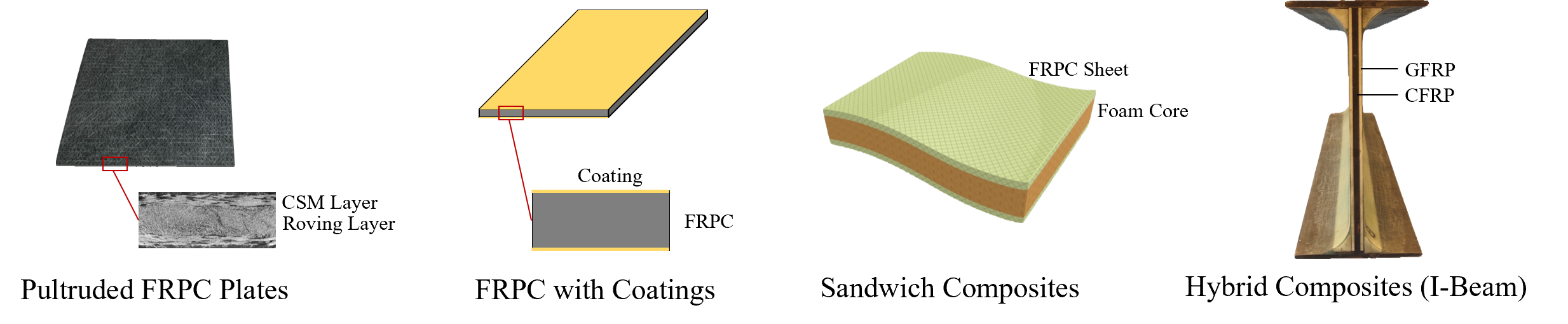}
     \caption{General application scenarios of multi-layered FRPC systems}
     \label{fig:examples_multilayer_composites}
 \end{figure}

For the practical design of FRPCs accounting for their potential degradation due to moisture ingression, it is crucial to establish a total or effective diffusion coefficient of such multi-layered systems. Fig~\ref{fig:examples_multilayer_composites} illustrates from left to right the different layers present in various multi-layered FRPCs, such as the chopped strand mat (CSM) layers and roving layers in pultruded FRPC plates, the FRPC plate with coating layers, the core and surface layers in composite sandwich structures, and a hybrid composite I-beam consisting of both GFRP and CFRP layers. In these layered composite structures, considering durability and resistance to water diffusion, it is essential to consider the design and optimization for the cross-section of multi-layered FRPCs. To ensure resistance to moisture diffusion, it is imperative to establish the relationship between the diffusion coefficient of a single layer and the total diffusion coefficient of the multi-layered materials. 

Numerous studies exist on the moisture diffusion of pultruded FRPC profiles \cite{Xin2016,Karbhari2009,Xin2021} and FRPC with coatings \cite{Newill1999,Barraza2003}. However, these studies treat them as homogenized materials without distinguishing between the different layers. Other researchers have investigated the diffusion behavior and proposed relevant models for generic multi-layered materials. Millers\cite{Miller1981} proposed a method for analyzing moisture absorption in multi-layered materials by adjusting the actual thickness of different layers into effective thicknesses and converting a multi-layered system into a single material system whose moisture diffusion can be established by the Fickian model. Avilés and Aguilar-Montero \cite{Aviles2010} investigated the moisture diffusion behavior of sandwich structures and their layers and found that the face sheet can effectively prevent moisture ingression into the sandwich structure. Nurge et al. \cite{Nurge2010} developed a finite difference method for describing the water uptake of multi-layered composites and sandwich structures. This method accurately predicted the moisture absorption rate of samples exposed to a fixed temperature and relative humidity after applying a mass conservation condition at the interface. Joshi and Muliana\cite{Joshi2010} presented an analytical solution for moisture absorption in sandwich structures based on the condition of continuous moisture concentration at the interface. Yu and Zhou\cite{Yu2018} also developed a diffusion model for the interface concentration problem, using the continuous normalized concentration and the conservation of mass condition. The proposed model's consistency was verified by finite element analysis and moisture absorption experiments. However, {\bf the studies mentioned above do not show a direct relationship between the individual layers' diffusion coefficient and the multi-layered materials' total diffusion coefficient.} These methods usually require iterative calculations to establish the concentration distribution within the material over time, which is very complex.

 \begin{figure}[h!]  
     \centering
     \includegraphics[width=1\linewidth]{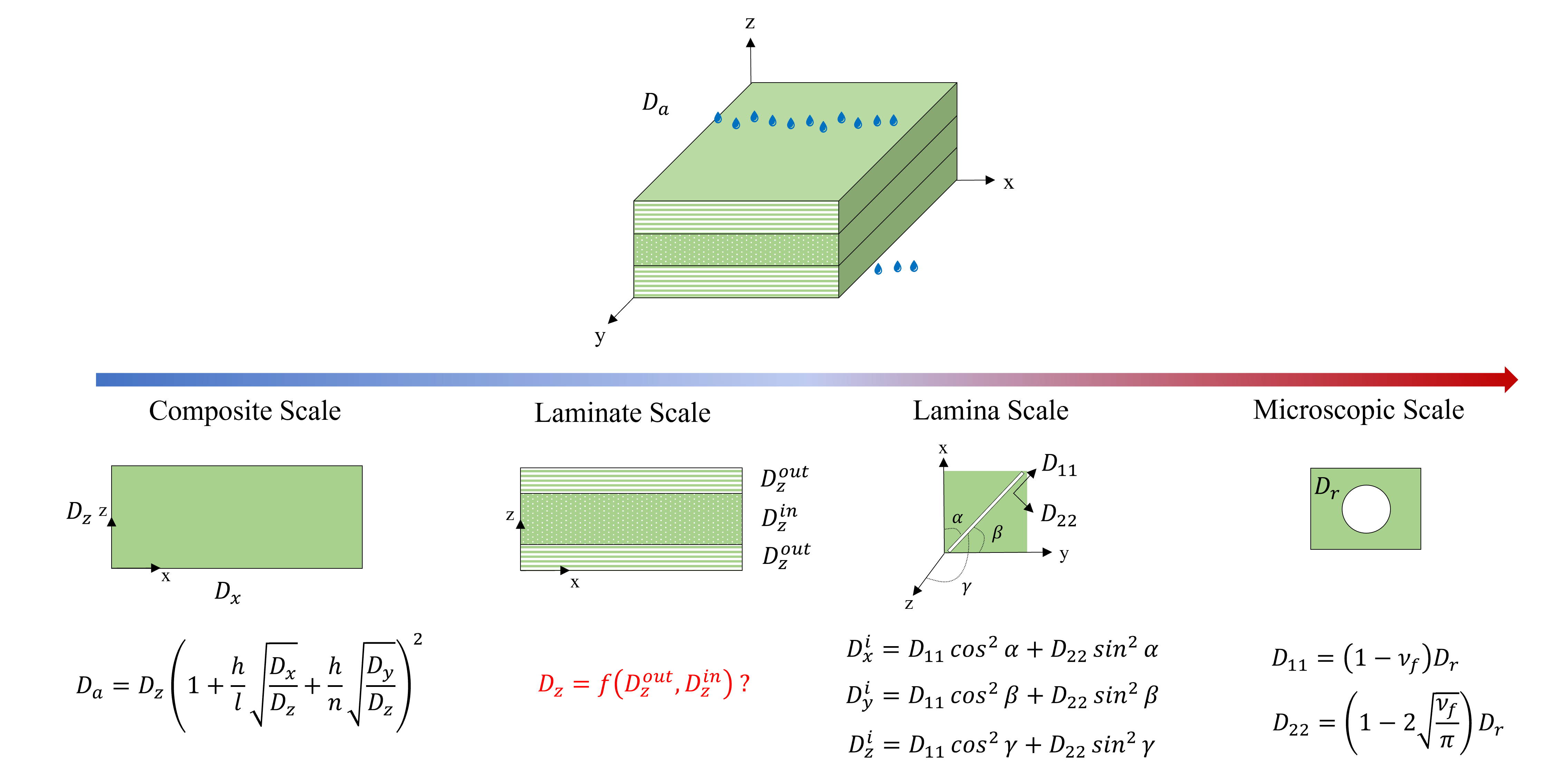}
     \caption{Description of moisture diffusion coefficients in multi-layered materials at different length scales}
     \label{fig:diffusion_scales}
 \end{figure}

Although each layer in multi-layered materials can fit the Fickian diffusion model, the overall moisture absorption behavior of multi-layered materials may exhibit a different behavior than Fickian diffusion. The scales of moisture diffusion in multi-layered materials can be categorized into four: composite scale, laminate scale, lamina (layer) scale, and microscopic scale, as shown in Fig~\ref{fig:diffusion_scales}. For the microscopic scale, the diffusion coefficients perpendicular ($D_{22}$) and parallel ($D_{11}$) to the fiber direction are given by Shen and Springer\cite{Shen1976} based on the analytical equation for heat conduction, as shown in Fig~\ref{fig:diffusion_scales}. Here, $v_f$ is the fiber volume fraction, and $D_r$ is the diffusion coefficient of the resin. The fiber is considered to be impermeable. At the lamina scale, the diffusion coefficients of the $i$th layer in the x-y-z coordinate system are provided based on fiber orientation\cite{Shen1976} defined by $\alpha$, $\beta$, and $\gamma$. $D_{11}$ and $D_{22}$ are the in-plane homogenized diffusion coefficients of a single layer, as mentioned above. Existing studies on moisture diffusion at the composite scale typically treat the layered composites as a single homogenized material during water immersion tests to determine the apparent diffusion coefficient ($D_a$) in the thickness direction. The relationship between $D_a$ and $D_x$, $D_y$, and $D_z$ can be obtained based on the edge effect equation\cite{Shen1976}, where $D_x$, $D_y$, and $D_z$ are the homogenized or total diffusion coefficients in the $x$, $y$, and $z$ direction of the composite, respectively. $l$, $n$, and $h$ are the entire composite's length, width, and thickness. In contrast, there are minimal relevant studies at the laminate scale between the composite and lamina scales, as highlighted in Fig~\ref{fig:diffusion_scales}. {\bf The relationship between the total diffusion coefficient, especially in the thickness direction ($D_z$) of multi-layered material systems, and the diffusion coefficient and other parameters of the individual layers or lamina is unclear.} 

The experimental weight gain measurements of multi-layered systems, including FRPCs, provide only the average or apparent moisture saturation value and apparent diffusion coefficient. Currently, there are two models for calculating the total diffusion coefficient of multi-layered FRPC laminates: the {\bf parallel and series models}. Shen and Springer \cite{Shen1976} first proposed the parallel model for the diffusion coefficient of composite laminates. Assuming that there are $N$ layers, the total diffusion coefficient of a multi-layered composite or laminate can be calculated by the parallel model for one-dimensional diffusion in a composite laminate. If the thickness of the $j${th} layer is $h_j$ ($j=1,2,3...N$) and the total thickness is $h$, for example, in the z-direction (thickness direction), $D_z$ is given by,

\begin{equation}\label{eq:parallel model}
    \centering
    D_z=\sum_{j=1}^N\frac{h_j D^j_{z}}{h}
\end{equation}

where $D^j_{z}$ is the diffusion coefficient of the $j$th layer. In contrast, the total diffusion coefficient can be also calculated by the series model as bellow\cite{Duncan2023},

\begin{equation}\label{eq:series model}
    \centering
    \frac{1}{D_z}=\sum_{j=1}^N\frac{h_j}{h D^j_{z}}
\end{equation}

These equations show that neither the parallel nor the series models account for the influence of the stacking order of layers. {\bf That is, the total diffusion coefficient cannot be identical for two multi-layered materials with the same volume fraction but different layer stacking orders.}  Rochowski and Pontrelli\cite {Rochowski2022} present a circuit-based model for studying mass diffusion in complex geometries, validating its accuracy against various solutions and emphasizing its utility in multi-layered systems. However, these studies are mainly applied in the medical or biological fields and have little relevance to the moisture diffusion of multi-layered FRPCs. Hence, the current paper examines the parameters that influence the total diffusion coefficient of multi-layered FRPCs.

This paper proposes a new unified model for determining the total diffusion coefficient of multi-layered materials. The moisture diffusion behavior of multi-layered materials is investigated through Finite Element Method (FEM) simulations and experiments that serve as input to the new models developed. Additionally, a method for designing and optimizing the cross-section (ratio of diffusion coefficient of different materials, percentage of individual materials chosen, and their stacking order) of multi-layered materials is proposed while considering resistance to moisture diffusion.


\section{Motivation}\label{motiv}

The existing parallel and series models to determine the total diffusion coefficient of multi-layered materials do not account for the stacking order of different layers in their expressions. This makes them unsuitable for designing and optimizing the cross-section of multi-layered materials to enhance their durability against moisture diffusion. Hence, we propose a new model for determining the total diffusion coefficient of multi-layered composites, accounting for several parameters that can influence their diffusion coefficient. To that end, a detailed parametric study of moisture diffusion through multi-layered materials is simulated using the Finite Element Method at the continuum length scale is performed. In addition, we conducted a water immersion test to verify the simulations.

Key innovations of this paper are as follows:

\begin{itemize}
    \item We aimed to investigate the fundamental reasons for the acceleration or deceleration of moisture diffusion in multi-layered materials dictated by the outer layer. Specifically, we focused on understanding how the surface layer affects moisture diffusion in layered or sandwich composites.
    \item We aimed to establish the effect of the ratio of diffusion coefficients, stacking order, and volume fraction of individual layers or lamina on the total diffusion coefficient of multi-layered materials.
    \item A new unified model for calculating the total diffusion coefficient of multi-layered materials is proposed, along with the application conditions.
    \item Additionally, a design methodology for composite structures with multi-layered materials considering long-term moisture resistance is provided.
\end{itemize}

Key assumptions in this work are that the composite material only undergoes one-dimensional Fickian diffusion and that the effect of the interface between different layers is not considered.


\section{Computational Modeling}\label{comp_method}
Computational modeling is performed using the Finite Element Method (FEM) to determine the total diffusion behavior of multi-layered materials. The key parameters considered in these models are discussed next, followed by details of the FEM simulation. 

\subsection{Parameters Considered}

Among several parameters affecting the moisture diffusion behavior of multi-layered composite systems, three critical parameters are considered in this paper - Ratio, Fraction, and Order. 

{\bf § Ratio of diffusion coefficients of individual layers (Ratio or R)}

In this paper, $D_{11}$ and $D_{22}$ are considered as in-plane moisture diffusion coefficients, and $D_{33}$ is the through-thickness coefficient. $D_{11}$, typically along the fiber direction, is generally larger than $D_{22}$ for continuous fiber-reinforced composites. $D_{33}$ is considered the same as $D_{22}$ in unidirectional composites, like in unidirectional pultruded FRPCs \cite{Xin2016,Xin2021}. But $D_{11}$ equals to $D_{22}$ in bi-directional woven FRPCs, and are different from $D_{33}$. In this paper, we only focus on the diffusion behavior of multi-layered composite systems in the $D_{11}$ and $D_{33}$ directions, which are assumed to be in the in-plane and through-thickness directions.

The moisture diffusion coefficients of several types of FRPCs with epoxy resin as the matrix are listed in Table \ref{tab:Diffusion coefficient of FRPC}. These coefficients differ by order of magnitude due to the different reinforcements, i.e., the diffusion coefficient of composites with flax (natural) fibers is higher than that with carbon (synthetic) fibers. In the FEM simulations discussed later, the value of $D_{33}$ is assumed to be $1\times 10^{-8} mm{^2}/s$ as a reference material (material 2), which is representative of materials similar to the epoxy/carbon fiber-based composites. {\bf The Ratio "R" is defined as the ratio of the diffusion coefficient of a different material (material 1) to the reference material (material 2)}. The range for R considered is from 1 to 100, with 10 and 100 representing materials similar to epoxy or epoxy/glass and epoxy/flax composites, respectively. Since our study individually involves 1D diffusion behavior in each orthogonal direction, we consider $D_{11}$ equal to $D_{33}$ for each material layer when performing the parametric analysis. 

\begin{table}[h!]
    \centering
        \caption{Moisture diffusion coefficients of common fiber-reinforced epoxy-based polymer composites}
    \begin{tabular}{|c|c|c|c|c|c|} \hline  
         Resin&  Fiber&  $\nu_f$&  Temperature&  $D_{33}$ ($mm^2/s$) & Reference\\ \hline  
         Epoxy&  Flax& $0.51$& Room & $10.5\times10^{-7}$ & M. Assarar et al. 2011\cite{Assarar2011}\\ \hline  
         Epoxy& Carbon& $0.69$& $23^\circ$C & $0.077\times10^{-7}$&Karbhari and Xian 2009\cite{Karbhari2009}\\ \hline  
         Epoxy& Glass& $0.33$& Room& $1.38\times10^{-7}$&Fan et al. 2019\cite{Fan2019}\\ \hline  
         Epoxy& - & - & Room& $1.41\times10^{-7}$&Fan et al. 2019\cite{Fan2019}\\ \hline 
    \end{tabular}
    \label{tab:Diffusion coefficient of FRPC}
\end{table}

{\bf § Volume fraction of individual layers (Fraction or F)}

The effect of the volume fraction of individual layers on the total diffusion coefficient of the multi-layered FRPCs is considered in both the series and parallel models. {\bf Fraction "F" represents the fraction of a particular material in the entire laminate}. Therefore, the volume fraction of material 1 in the multi-layered composites model is set to $0, 0.2, 0.4, 0.6, 0.8$, and $1$ in the FEM simulations. Here, $0$ implies that there is no material 1 layer in the entire laminate and has only material 2. On the other hand, $1$ means that the entire laminate only has material 1.

{\bf § Stacking order of individual layers (Order) }

\begin{figure}[h!]
    \centering
    \includegraphics[width=0.3\linewidth]{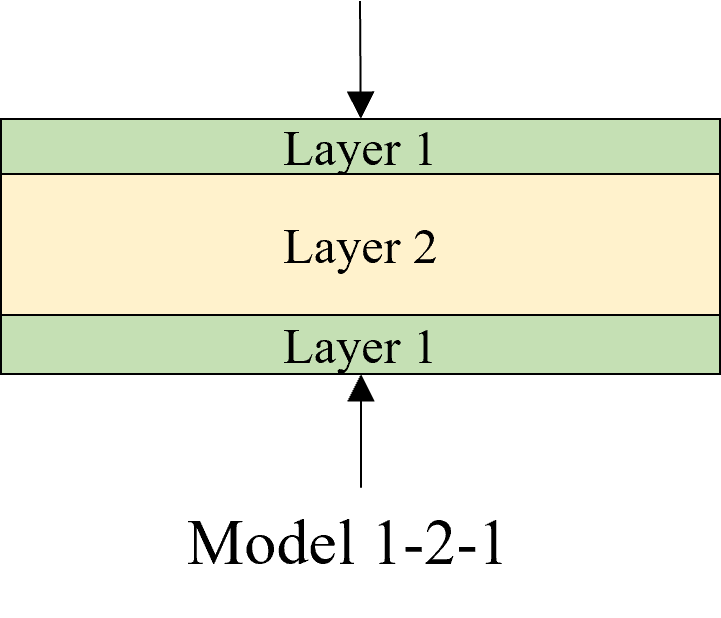}
    ~ ~
    \includegraphics[width=0.3\linewidth]{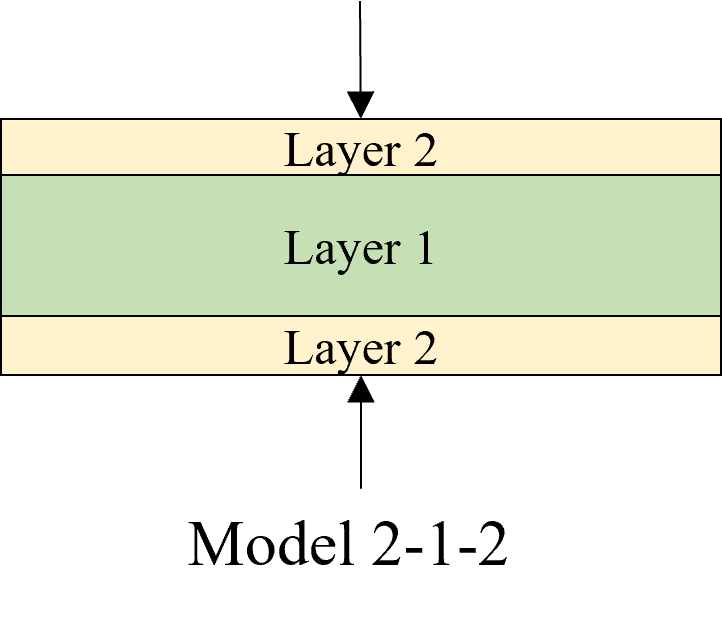}
    \caption{Cross-section of Model 1-2-1 and Model 2-1-2 displaying their layer order}
    \label{fig:two models}
\end{figure}

As the number of layers and material types increases, the stacking order becomes more complex. We consider multi-layered composite systems with only three layers and two materials. Keeping the stacking symmetric through the thickness, the outer layers (top and bottom layers) are assigned the same material as it is relevant for most practical situations, as shown in Fig~\ref{fig:examples_multilayer_composites}. The layer with material 1, which has the larger diffusion coefficient, is defined as layer 1, and that with material 2 is described as layer 2. Therefore, there are two types of models in the finite element simulation, Model 1-2-1 and Model 2-1-2, as shown in Fig~\ref{fig:two models}. All the parametric studies reported in their paper are performed for these two models and will be discussed separately.

\subsection{Finite Element Method Simulations}
\begin{figure}[h!]
    \centering
    \includegraphics[width=0.5\linewidth]{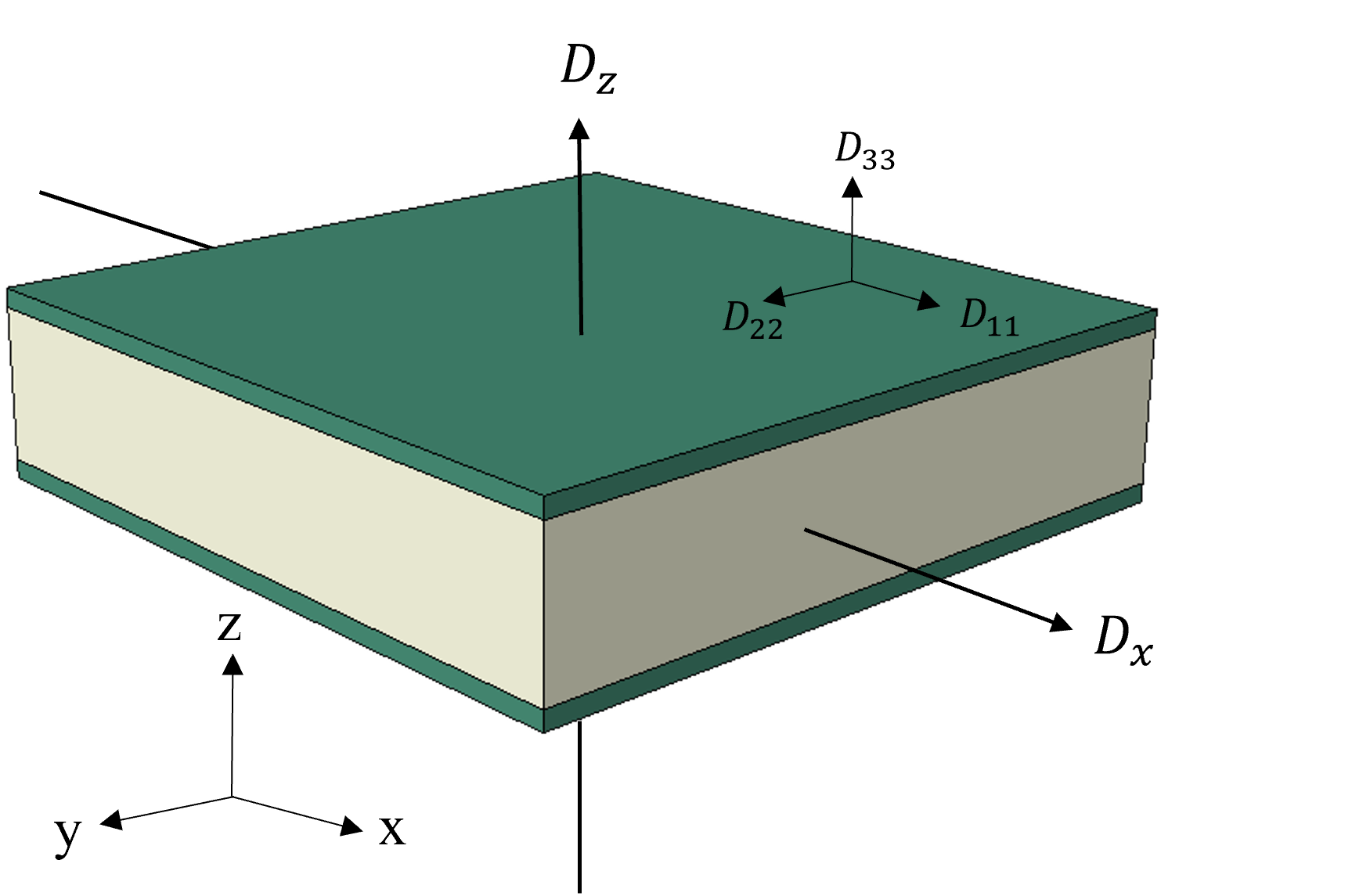}
    \caption{Finite element method model of moisture diffusion in Model 1-2-1 in the $x$ and $z$ directions}
    \label{fig:one direction FEM}
\end{figure}

The multi-layered composite system is modeled using a mass diffusion module in the Finite Element Method software, Abaqus 2020. 8-node linear heat transfer brick elements DC3D8 are used for meshing the geometry. All models have a rectangular geometry, with dimensions of 40 mm $\times$40 mm $\times$ 10 mm, and contain three parallel layers in the thickness direction and two types of materials. As mentioned above, layer 1 and layer 2 correspond to materials 1 and 2, respectively. The moisture diffusion of multi-layered composites is assumed to be one dimensional in each of the two perpendicular directions $x$ and $z$, corresponding to $D_{11}$ and $D_{33}$ of each layer, respectively, as shown in Fig~\ref{fig:one direction FEM}. The mesh refinement is considered only in the diffusion direction to minimize the computational effort due to one-dimensional diffusion consideration. A mesh sensitivity analysis is performed so that further mesh refinement does not influence final results. The mesh size was set to $0.25$ mm in the diffusion direction.

The governing equation for transient moisture diffusion is given by,

\begin{equation}\label{eq:governing equation}
    \centering
    \begin{split}
    \frac{\partial c}{\partial t}+\nabla \cdot J =0 \\
    J= - D \nabla c
    \end{split}
\end{equation}
where $c$ is the moisture concentration, $J$ is the concentration flux, and $D$ is the diffusion coefficient. In Abaqus, the concentration is normalized generally when applying mass diffusion for multi-layered materials because of the interfacial heterogeneity,

\begin{equation}\label{eq:interfacial consideration}
    \centering
    \begin{split}
    \phi = c / s\\
    J=-D [s\frac{\partial \phi}{\partial x}+\phi \frac{\partial s}{\partial x}]
    \end{split}
\end{equation}
where $s$ is the solubility. Generally, solubility is a constant for each material in composites, so Eq~\ref{eq:interfacial consideration} reduces to, 

\begin{equation}\label{eq:SOL defination}
    \centering
    J=-D s\frac{\partial \phi}{\partial x}
\end{equation}
In Abaqus, $SOL$ is the total amount of solute in the model calculated as the sum of the product of the mass concentration and the integration point volume over all elements. For a model with $n$ types of materials, $SOL$ is calculated by, 

\begin{equation}\label{eq:SOL calculation}
    \centering
    SOL=\sum _{j=1} ^{n} \sum _{i=1} ^{m_j} c_{ji} V_{ji}
\end{equation}
where $m_j$ is the number of integration points in the $j$th material, $c_{ji}$  and $V_{ji}$ are the mass concentration and volume of the $i$th integration point in the $j$th material, respectively. For multi-material composites with $n$ types of materials, $s_j$, $V_j$, and $\rho_j$ are the solubility, volume, and density of the $j$th material, respectively. The moisture gain in composites at time $t$ is given by,

\begin{equation}\label{eq:Mt calculated from SOL}
    \centering
    M_t  = \frac{\rho _w \sum _{j=1} ^{n} \sum _{i=1} ^{m_j} c_{ji} V_{ji}} {\sum _{j=1} ^{n} \rho_j V_j} 
         = \frac{\rho _w \; SOL}{\sum _{j=1} ^{n} \rho_j V_j}
\end{equation}
where $\rho _w$ ($1000 \ kg/m^3$) is the density of water. It should be noted that the solubility of individual material is related to the saturation concentration as,
\begin{equation}\label{eq:scocm}
    \centering
    c_m^j=c_0\times s_j
\end{equation}
where, $c_0$ and $c_m^j$ are moisture concentration at the boundary and the saturation concentration of the $j$th material, respectively. Combining Eq~\ref{eq:interfacial consideration} and Eq~\ref{eq:scocm}, the normalized mass concentration of the $i$th integration point in the $j$th material of the multi-layered model is given by,

\begin{equation}\label{eq:explanation phi}
    \centering
    \phi_{ji}= \frac{c_{ji}}{s_j}=\frac{c_{ji}}{c^j_m}c_0 
\end{equation}
Generally, the boundary condition of the two sides exposed to moisture in FEM simulation is assumed to be $\phi=c_0$ ($c_{ji}=c^j_m$), which means material is saturated at the boundary. So, the moisture gain in composites at saturation is given by, 

\begin{equation}\label{eq:Mm calculated from SOL}
    \centering
    M_m  = \frac{\rho _w \sum _{j=1} ^{n} \sum _{i=1} ^{m_j} c_m^j V_{ji}}{\sum _{j=1} ^{n} \rho_j V_j} = \frac{c_0\rho _w \sum _{j=1} ^{n} s_j V_j}{\sum _{j=1} ^{n} \rho_j V_j}
\end{equation}
The normalized moisture gain is calculated using Eq~\ref{eq:Mt calculated from SOL} and Eq~\ref{eq:Mm calculated from SOL} as,

\begin{equation}\label{eq:Mt/Mm calculated from SOL}
    \centering
    \frac{M_t}{M_m}= \frac{SOL}{c_0\sum _{j=1} ^{n} s_j V_j}
\end{equation}

Specially, for individual material where $n=1$ and composites with two types of material where $n=2$, the normalized weight gain is given by Eq~\ref{eq:Mt/Mm n=1 or 2}, 

\begin{equation}\label{eq:Mt/Mm n=1 or 2}
    \centering
    \frac{M_t}{M_m}= \frac{SOL}{c_0s V} ~  (n=1)  \quad \; \mathrm{and} \quad \; \frac{M_t}{M_m}= \frac{SOL}{c_0(s_1 V_1 + s_1 V_2)} ~  (n=2)
\end{equation}

Since $SOL$ of a model can be extracted directly as an output in Abaqus 2020, the moisture gain of individual material and multi-layered composites can be calculated using Eq~\ref{eq:Mt/Mm calculated from SOL} in this paper. Although the different solubility of two types of material can affect the final moisture gain indeed, which we will discuss in Section~\ref{effect_of_solubility}, the solubility of most types of FRPCs is not significantly different. So, we assume $s=s_1=s_2$ ($c_m=c^1_m=c^2_m$) for simplification in the parametric analysis by FEM simulation. In this case, the weight gain of individual material and multi-layered composites in parametric analysis is given by, 

\begin{equation}\label{eq:Simplified Mt/Mm calculated from SOL}
    \centering
    \frac{M_t}{M_m}= \frac{SOL}{c_0sV} ~ (Simplified)
\end{equation}

All of $s$ in the FEM parametric analysis can be set to be $1$. For convenience, we can also set $c_0=1$, in which case $\phi=1$ at the boundary. A total of $146$ models with different input parameters are used for the parametric study, of which $97$ model diffusion in the $z$ direction, and the other $49$ are in the $x$ direction.

From the data derived from the FEM simulations, the initial $60\%$ of each response data adheres to a linear correlation if it follows the Fickian diffusion model, shown in Eq~\ref{eq:linear model}\cite{Assarar2011},

\begin{equation}\label{eq:linear model}
    \centering
    \frac{M_t}{M_m}=\frac{4}{h}\sqrt{\frac{D \: t}{\pi}}
\end{equation}

where $D$ is the total diffusion coefficient and $h$ is the thickness in the diffusion direction of the multi-layered composite model. The diffusion coefficient can be deduced via linear regression analysis. Subsequently, the entire response data of FEM simulation is compared with the theoretical approximation from the one-dimensional Fickian diffusion model\cite{Shen1976}, expressed as Eq~\ref{eq:approximation model},

\begin{equation}\label{eq:approximation model}
    \centering
    \frac{M_t}{M_m}=1-\exp[-7.3(\frac{D \: t}{h^2})^{0.75}]
\end{equation}

This approach ensures a comprehensive analysis, utilizing the theoretical foundation provided by the Fickian model to validate the numerical simulations conducted through the Finite Element Method.

\section{Experimental Work on Multi-Layered Pultruded Plate}\label{exp_method}

\begin{figure}[h!]
    \centering
    \includegraphics[width=0.8\linewidth]{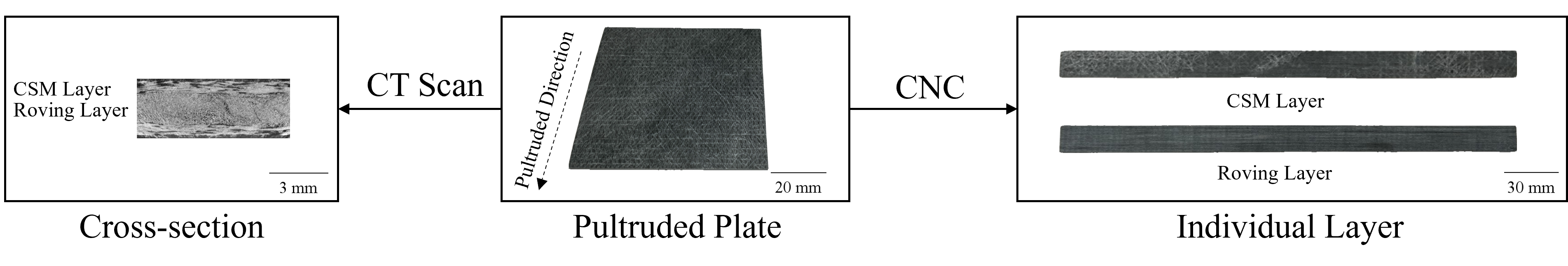}
    \caption{Details of the multi-layered pultruded plate (Model 1-2-1)}
    \label{fig:pultruded plate}
\end{figure}

The water immersion experiment under room temperature is conducted for Model 1-2-1 using a pultruded glass fiber-reinforced polyester composite plate manufactured by Spare Inc., China. This pultruded plate has two outer layers of chopped strand mat (CSM) and one inner layer of continuous roving, representing Model 1-2-1. The roving fiber is an E-glass fiber named 386H, provided by China Jushi Co., Ltd., China, and the resin is a m-phenylene resin named EL-400, provided by Zhenjiang Leader Composite Co., Ltd, China. The CSM is a fiberglass stitched mat named EMK300 provided by Changzhou Zhongjie Composites Co., Ltd., China. Individual CSM and roving layers are extracted from the whole plate by the CNC process and are shown in Fig~\ref{fig:pultruded plate}.  Details of the extracted layer specimens are summarized in Table \ref{tab:specimens of multil-layer}. Chopped fibers are oriented randomly in-plane in the CSM layer such that $D_{11} = D_{22}$, but different from $D_{33}$. For the unidirectional roving layer, $D_{22} = D_{33}$, but different from $D_{11}$ because the roving fibers are oriented along the pultruded direction. The focus is on $D_{33}$ of these two individual layers whose length or width-to-thickness ratio is high enough to neglect the edge effect. The moisture gain is calculated using Eq~\ref{eq:weight gain calculation} by weighing samples periodically,

\begin{equation}\label{eq:weight gain calculation}
    \centering
    M_t= \frac{m_t-m_0}{m_0}
\end{equation}

where $m_t$ and $m_0$ is the weight measured of the sample when time is $t$ and $0$, respectively.  

\begin{table}[h!]
    \centering
        \caption{Specimens specification of multi-layered pultruded composite plate used for water immersion test}
    \begin{tabular}{|c|c|c|c|c|} \hline  
         Specimen&  Material& Length ($mm$)&  Width ($mm$)&  Thickness ($mm$)\\  \hline
         CSM layer&  Resin, chopper fiber& \multirow{2}{*}{$250$} & \multirow{2}{*}{$15$}& $0.95$\\  \cline{1-2} \cline{5-5} 
         Roving layer& Resin, roving fiber& & & $0.87$\\ \hline 
         Pultruded plate& Resin, chopper fiber, and roving fiber& $50$& $50$& $3.13$\\\hline

         \end{tabular}
    \label{tab:specimens of multil-layer}
\end{table}

The details of each layer within the pultruded plate used for FEM geometry are shown in Table \ref{tab:details of each layer in pultruded plate}. It is assumed that the inner layer is in the middle of the multi-layered material, which is symmetric about the mid-plane. Based on the experimental results, FEM simulations of actual specimens will be conducted to verify the accuracy and reasonableness of the parametric simulation analysis above. It should be mentioned that $M_t$ of multi-layered pultruded plate is calculated using Eq~\ref{eq:Mt/Mm n=1 or 2} ($n=2$) since the solubility of individual material is different.

\begin{table}[h!]
    \centering
    \caption{Details of each layer in the multi-layered (three-layered) composite material considered for experiments}
    \begin{tabular}{|c|c|c|c|c|} \hline  
        \multirow{2}{*}{Specimen} & \multicolumn{2}{c|}{Outer layer} & \multicolumn{2}{c|}{Inner layer} \\ \cline{2-5}
        & Material & Thickness ($mm$) & Material & Thickness ($mm$) \\ \hline 
        Pultruded plate (Model 1-2-1)& CSM layer & $0.74$ & Roving layer & $1.65$ \\ \hline 
    \end{tabular}
    \label{tab:details of each layer in pultruded plate}
\end{table}

\section{Results and Discussion}\label{resdis}

The weight gain responses from the Finite Element Method simulations of moisture diffusion through multi-layered composites (laminates) with different stacking orders and layer volume fractions are presented and discussed. In particular, moisture diffusion along the $x$ and $z$ directions is examined in detail.

\subsection{Simulation of Moisture Diffusion in the $x$ Direction ($D_{x}$)}

Considering the same in-plane diffusion in the $x$ and $y$ directions, we only focus on the $x$ direction. The results and conclusions of diffusion in the $x$ direction can also be applied in the $y$ direction. For in-plane diffusion, the stacking order of Model 1-2-1 and Model 2-1-2 does not influence the diffusion process. Hence, we only show the results of simulation from Model 1-2-1 in the $x$ direction here in Fig~\ref{fig:FEM result D11 model 1-2-1}.

Linear regression analysis of the simulation output data from Model 1-2-1, according to Eq~\ref{eq:linear model}, shows that the coefficient of determination ($R^2$) of all data is higher than $0.98$, which means that all data from Model 1-2-1 in the $x$ direction exhibit Fickian behavior (Fig~\ref{fig:FEM result D11 model 1-2-1}). It can be seen that the moisture gain always conforms to the Fickian diffusion model regardless of R and F change. It should be noted that the simulation data is higher than the Fickian diffusion model at the early stage of the diffusion process, while it is reversed at the later stage. This difference is decided by Eq~\ref{eq:approximation model}, which is only an approximate curve of the Fickian diffusion model.

\begin{figure}[h!] 
  \centering
  \subfigure[]
  {
      \label{fig:D11 comparsion}\includegraphics[width=0.48\linewidth]{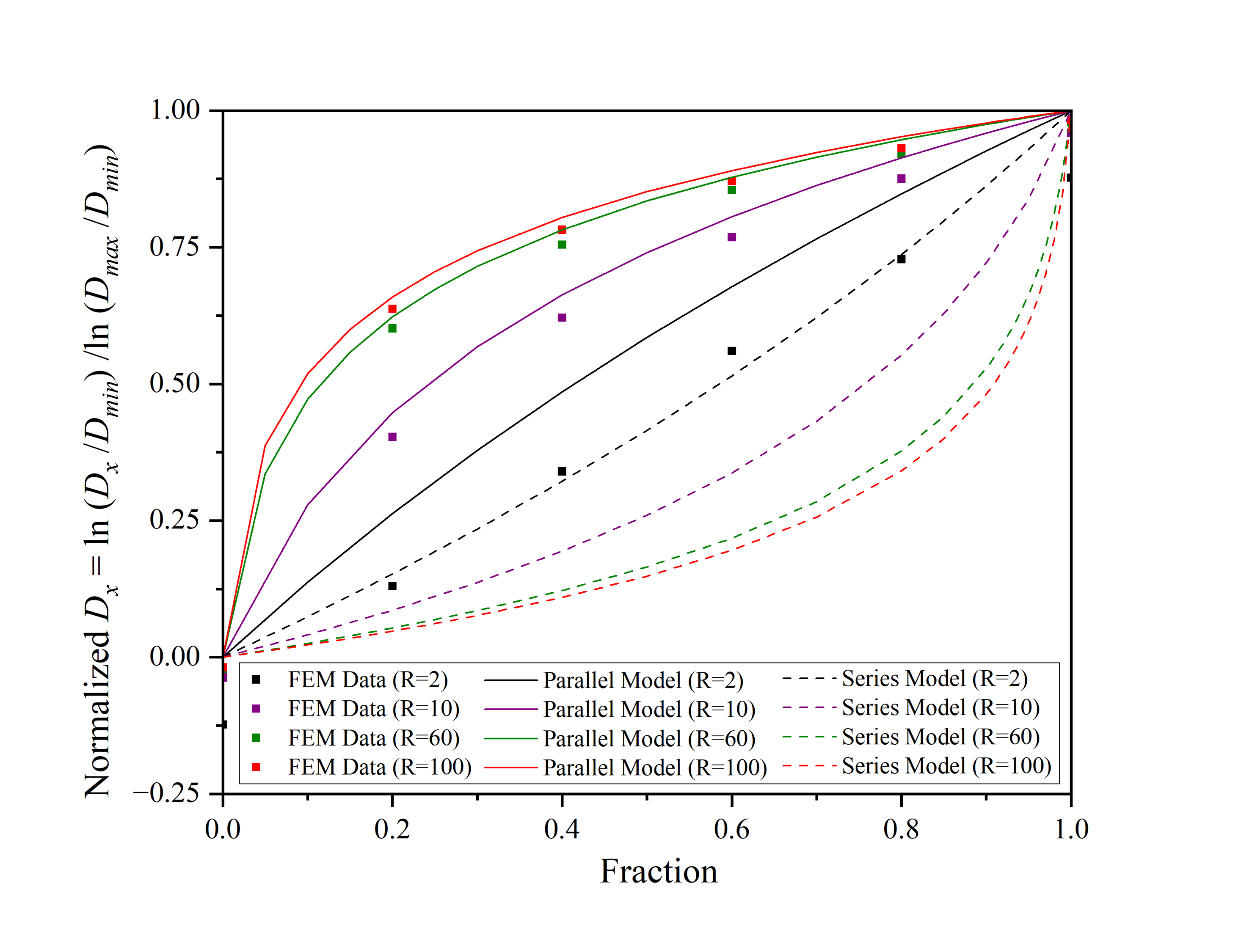}
  }
  \subfigure[]
  {
      \label{fig:heat map D11 model 1-2-1}\includegraphics[width=0.48\linewidth]{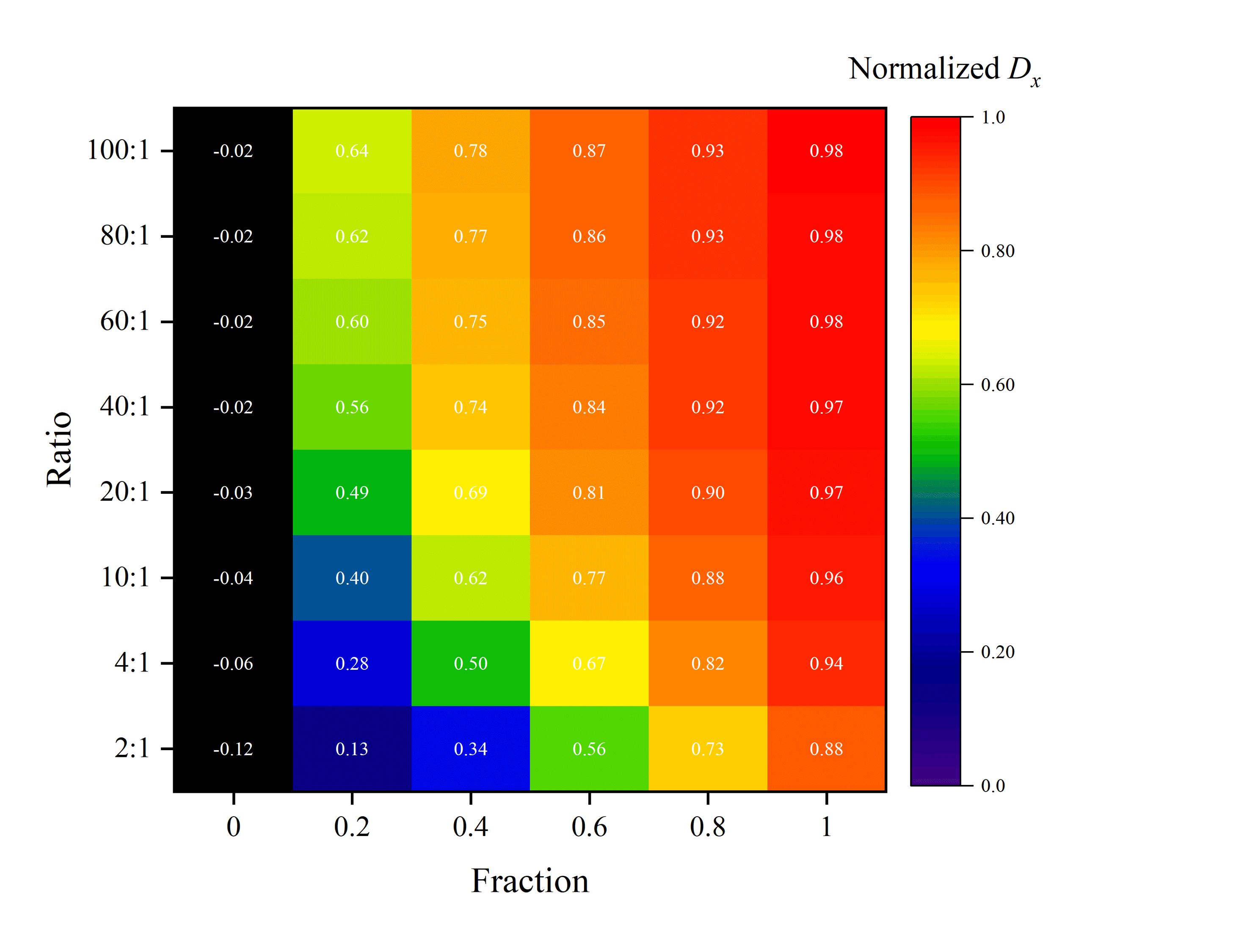}
  }
  
  \caption{ (a) Normalized $D_{x}$ of Model 1-2-1 in $x$ direction along with predicted value by the parallel and series model (b) Heat map of normalized $D_{x}$ with respect to R and F from Model 1-2-1.}
  \label{fig:parallel and series model D11 combined} 
\end{figure}

To eliminate the influence of R, the ratio of diffusion coefficients of individual layers, on the magnitude of diffusion coefficients, we implement a normalization procedure for the diffusion coefficient $D$ determined through data fitting, as shown in Eq~\ref{eq:normalized D by min and max}, 

\begin{equation}\label{eq:normalized D by min and max}
  Normalized \: D = \frac{\ln{⁡(D/D_{min})}}{\ln⁡{(D_{max}/D_{min})}}
\end{equation}

where $D_{max}$ and $D_{min}$  refer to  $D^1_{11}$ and $D^2_{11}$, which are the diffusion coefficients in the $x$ direction of layer 1 and layer 2, respectively. Fig~\ref{fig:D11 comparsion} shows the calculated total diffusion coefficients according to Eq~\ref{eq:linear model}, along with the predicted total diffusion coefficients of the parallel model and series model according to Eq~\ref{eq:parallel model} and Eq~\ref{eq:series model}, for Model 1-2-1 corresponding to R values of $2$, $10$, $60$, and $100$. The vertical axis is normalized $D_{x}$, where $D_{x}$ is the total diffusion coefficient of the multi-layered composites. For R of $2$, both the parallel model and series model can be used to calculate the total diffusion coefficient of Model 1-2-1, as the two models yield very close responses. The normalized $D_{x}$ of Model 1-2-1 is closer to the parallel model for R of $10$, $60$, and $100$ compared to that for R $= 2$. That is, when the ratio of the diffusion coefficients of individual layers in a multi-layered composite is smaller, the result of parallel and series models is similar. 

Overall, it should also be noted that the calculated total diffusion coefficient from the FEM simulations is slightly lower than those predicted by the parallel model. This deviation is introduced by Eq~\ref{eq:linear model}, which implies that the diffusion coefficient calculated by the linear model inherently is smaller than the diffusion coefficient from the FEM simulations. This method is widely used to calculate the moisture diffusion of composites \cite{Assarar2011,Fan2019,JIANG2014,JOLIFF2012}, making this deviation acceptable in this paper.  

In summary, the total diffusion coefficient of Model 1-2-1 in the $x$ direction, in-plane direction, can be predicted by the parallel model and not the series model. Only when R is small, and approaches 1, that is, a single material from a diffusion perspective, the total diffusion coefficient of the multi-layered composites is close to both the parallel and the series model.

The heat map of normalized $D_{x}$ with respect to F and R variables is shown in Fig~\ref{fig:heat map D11 model 1-2-1}.
It can be seen that the total diffusion coefficient of Model 1-2-1 in the $x$ direction depends on both F and R, increasing with both F and R increasing. It is important to highlight that the occurrence of values below $0$ in Fig~\ref{fig:heat map D11 model 1-2-1} are also attributed to the calculation accuracy of Eq~\ref{eq:linear model}. These negative data points can be considered to be $0$, which means that $D_x$ is very close to $D_{min}$.

\subsection{Simulation of Moisture Diffusion in the $z$ Direction ($D_{z}$)}

\subsubsection{Model 1-2-1}

\begin{figure}[h!] 
  \centering
  \subfigure[]
  {
      \label{fig:3D Map D33 model 1-2-1}\includegraphics[width=0.48\linewidth]{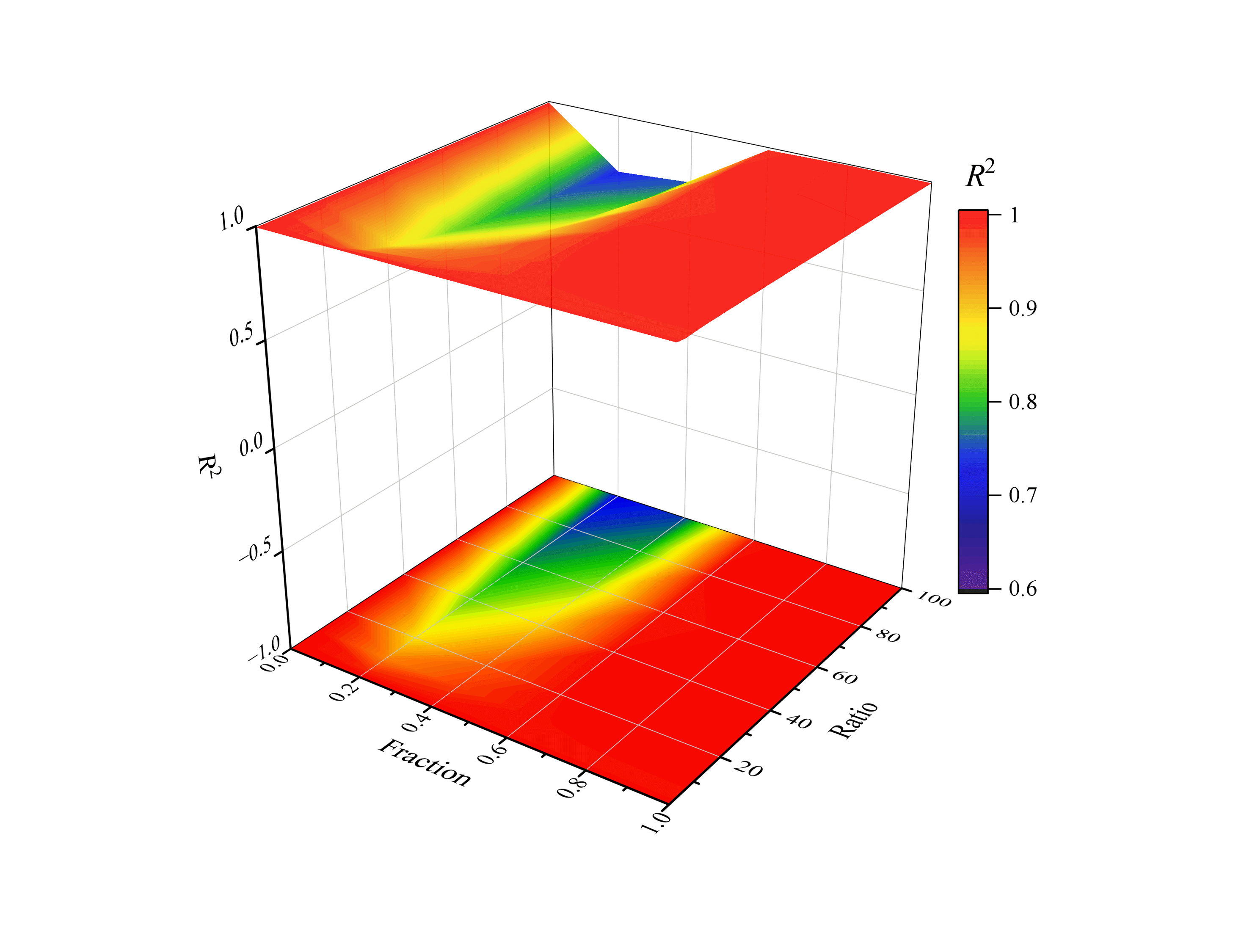}
  }
  \subfigure[]
  {
      \label{fig:Boundary map D33 model 1-2-1}\includegraphics[width=0.48\linewidth]{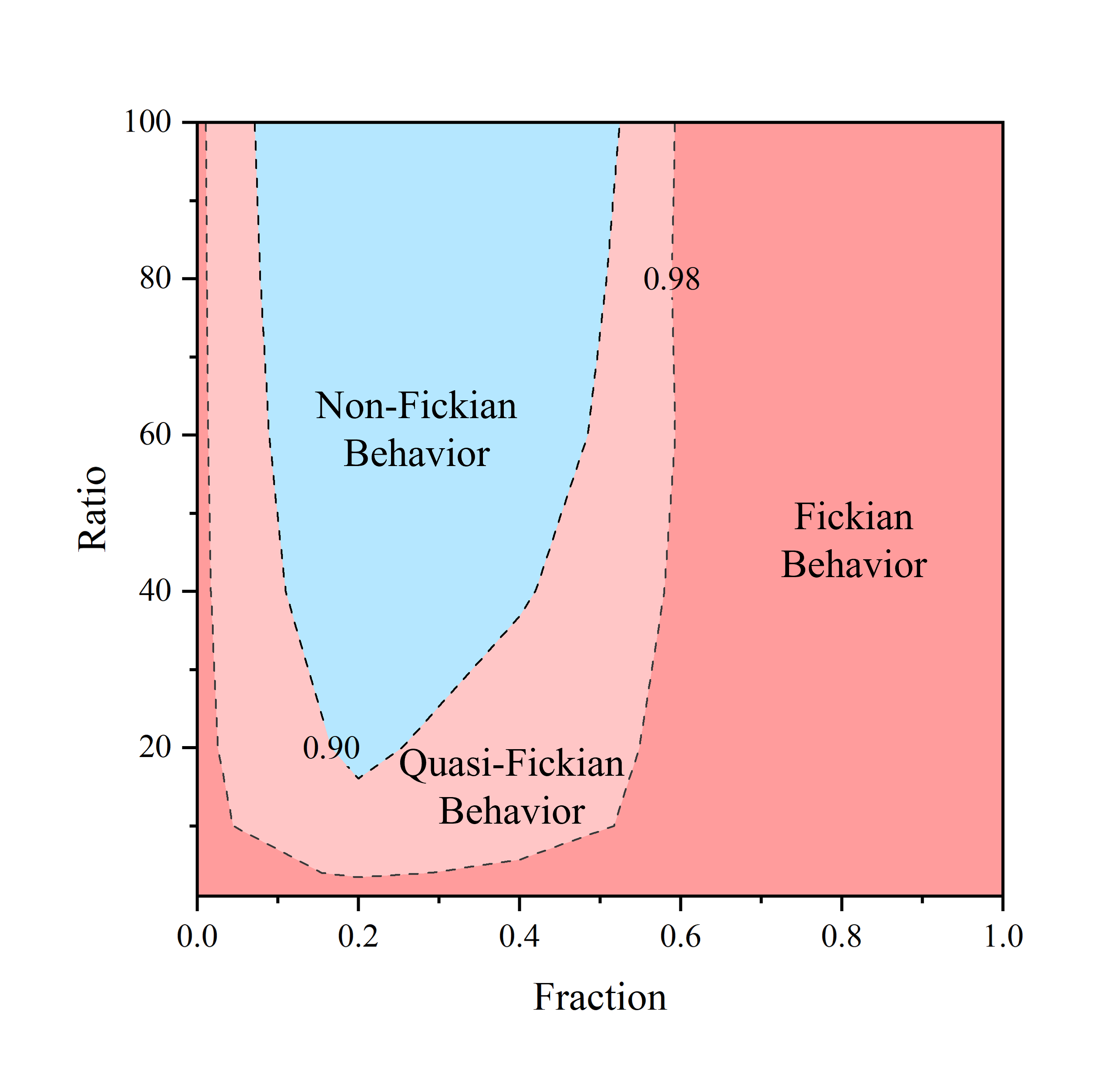}
  }
  \caption{Results of linear regression analysis of the simulation data from Model 1-2-1 in $z$ direction (a) Three dimension contour map of $R^2$ with respect to R and F (b) Boundary map of $R^2$ with respect to R and F indicating clear regions of Fickian and non-Fickian behavior with a band of quasi-Fickian behavior.}
  \label{fig:R2 model 1-2-1} 
\end{figure}

Linear regression analysis of the FEM simulation output data from Model 1-2-1, according to Eq~\ref{eq:linear model}, produces a three-dimensional contour map illustrating the coefficient of determination ($R^2$) with respect to the F and R variables, as shown in Fig~\ref{fig:R2 model 1-2-1}. A higher $R^2$ value indicates a better fit and a closer approximation to the Fickian diffusion model. Fig~\ref{fig:Boundary map D33 model 1-2-1} also shows a discernible regularity and symmetry in $R^2$ across the F and R axes. Specifically, in regions where $R^2$ exceeds $0.98$ (deep pink region), it is inferred that the multi-layered composite system mainly adheres to Fickian behavior, allowing the application of the one-dimensional Fickian model to calculate the total diffusion coefficient with high accuracy. Conversely, in regions where $R^2$ falls below $0.9$ (blue region), the multi-layered composite system is considered to deviate from Fickian behavior, although the individual layers adhere to the Fickian diffusion principle. For intermediate values, $0.9 < R^2 < 0.95$, the multi-layered composite system is interpreted as exhibiting quasi-Fickian behavior. In such cases, while the one-dimensional Fickian model remains applicable, it is expected to introduce certain inaccuracies.

\begin{figure}[H] 
  \centering
  \subfigure[]
  {
      \label{fig:D33 model 1-2-1 R=2}\includegraphics[width=0.48\linewidth]{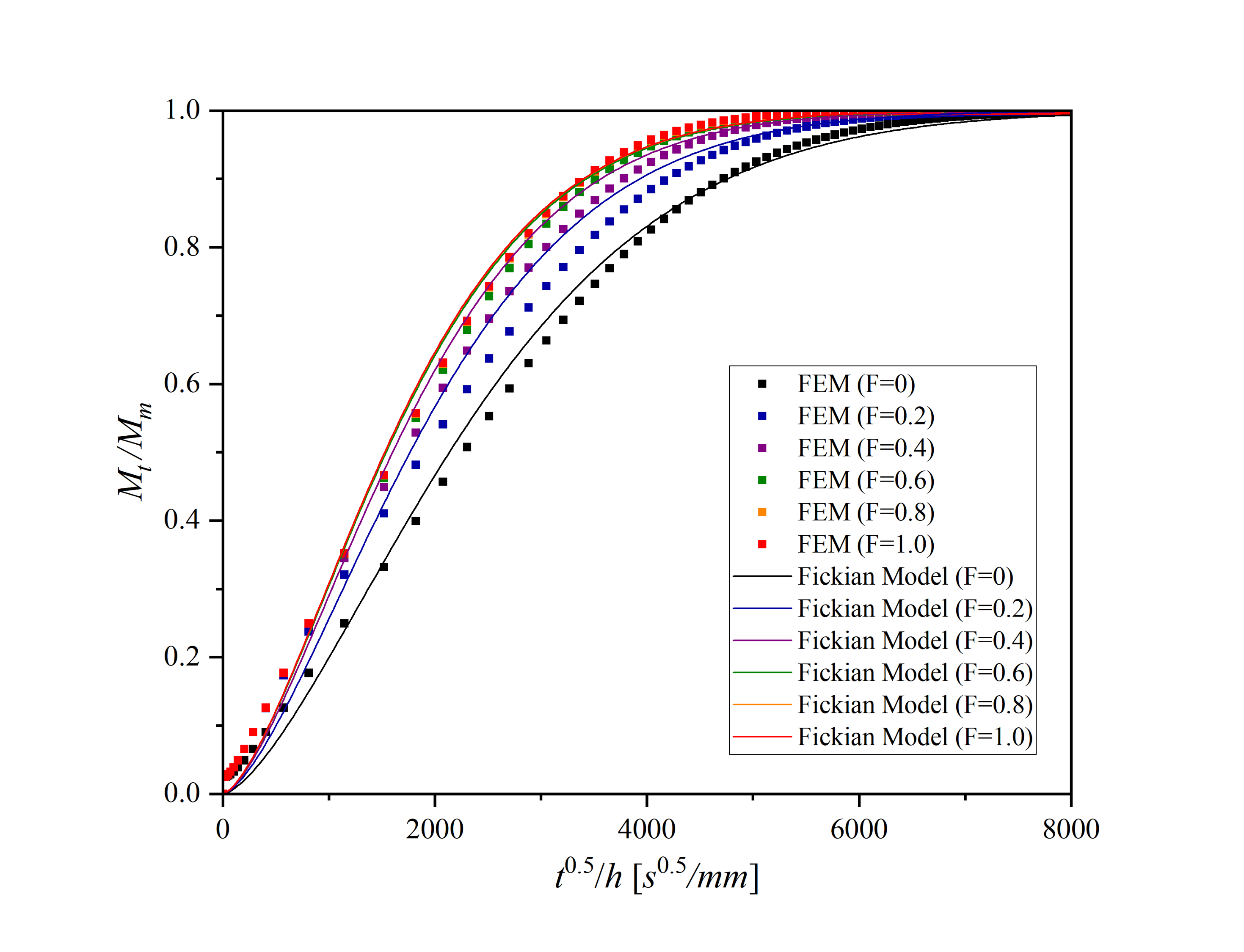}
  }
  \subfigure[]
  {
      \label{fig:D33 model 1-2-1 R=10}\includegraphics[width=0.48\linewidth]{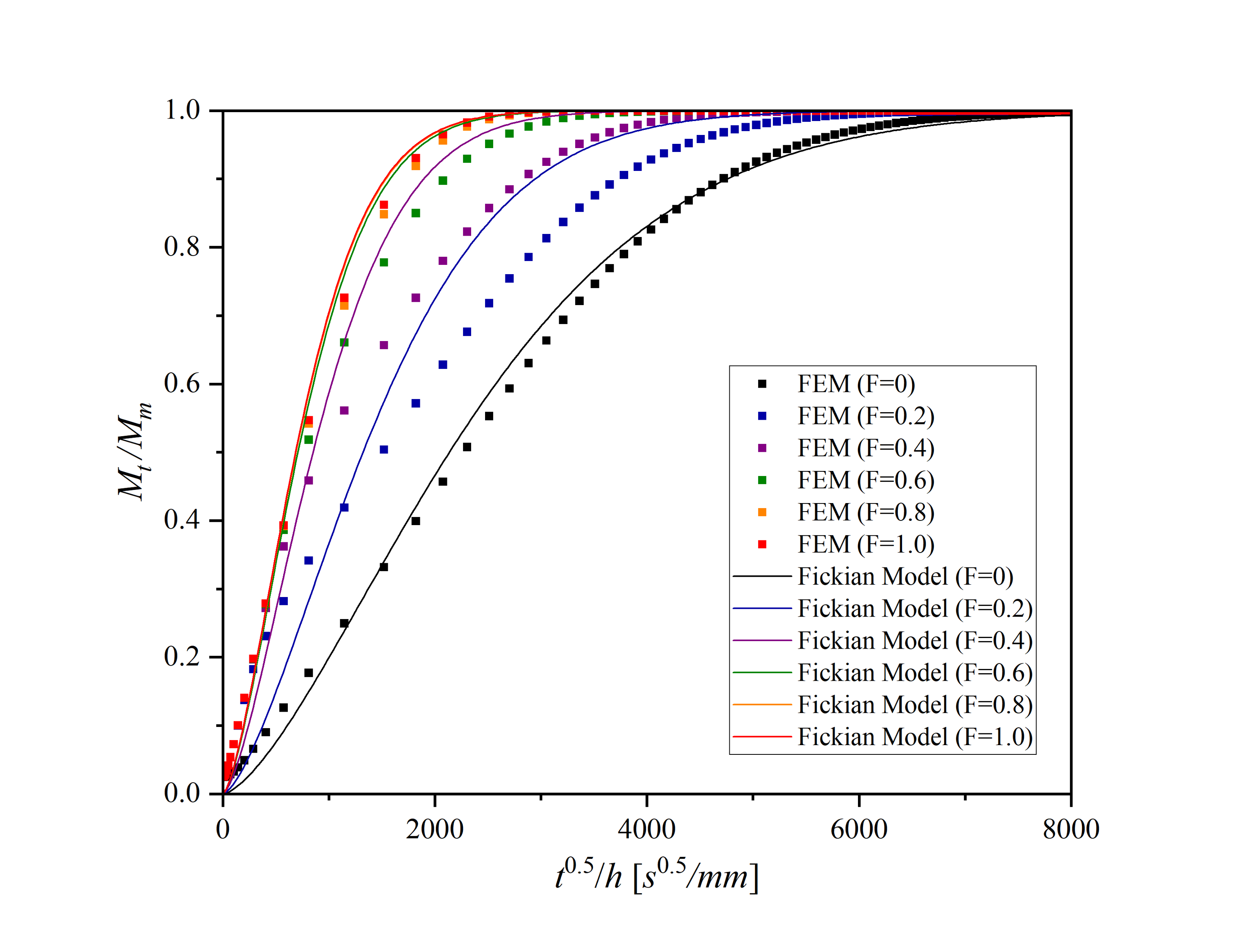}
  }
    \subfigure[]
  {
      \label{fig:D33 model 1-2-1 R=60}\includegraphics[width=0.48\linewidth]{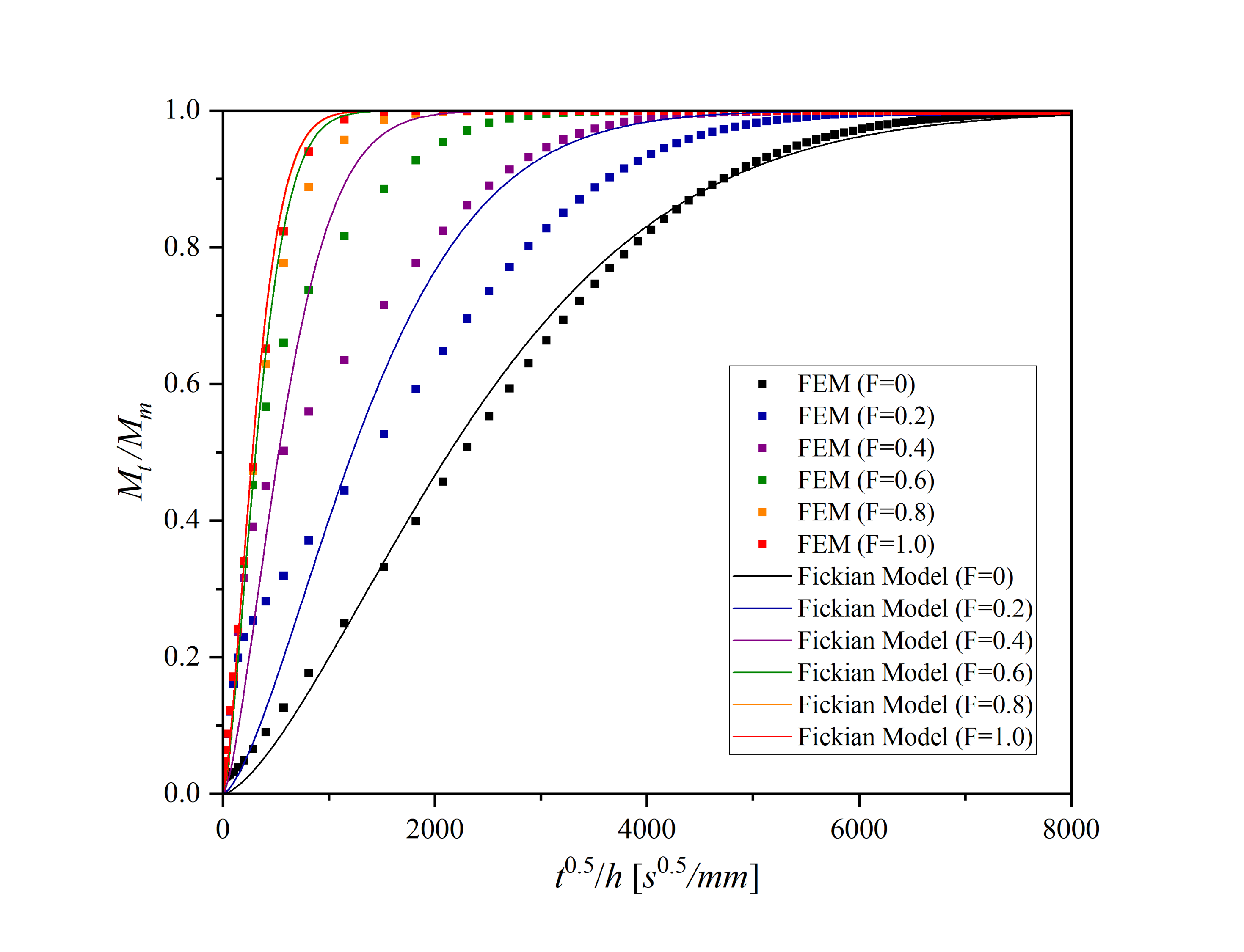}
  }
  \subfigure[]
  {
      \label{fig:D33 model 1-2-1 R=100}\includegraphics[width=0.48\linewidth]{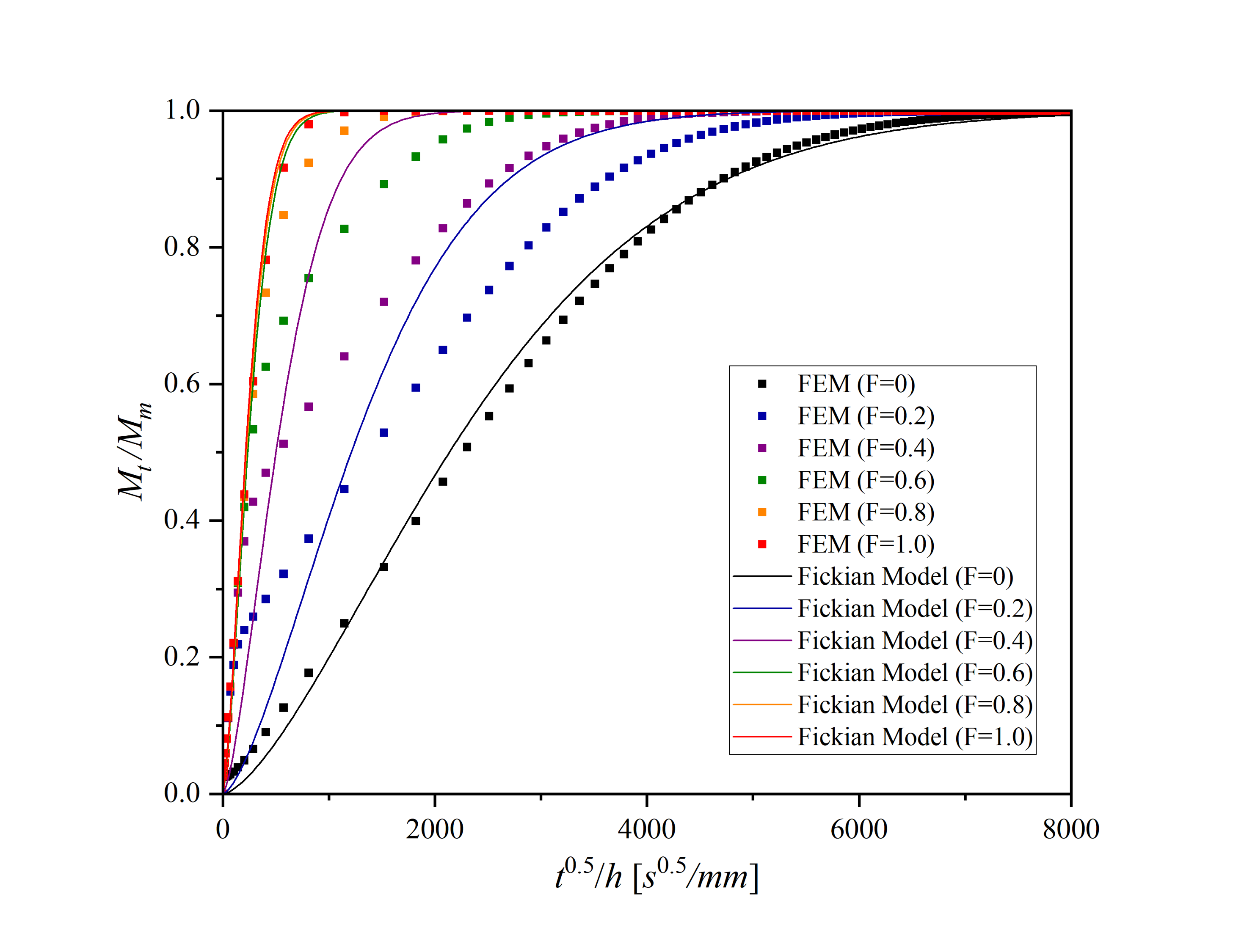}
  }
  \caption{ Results of simulation data and Fickian equation fit for Model 1-2-1 in $z$ direction (a) R=2 (b) R=10 (c) R=60 (d) R=100.}
  \label{fig:FEM result D33 model 1-2-1} 
\end{figure}

Fig~\ref{fig:FEM result D33 model 1-2-1} presents the result of Model 1-2-1 for R of $2$, $10$, $60$, and $100$. For R of $2$, the diffusion process adheres to the Fickian diffusion model, irrespective of variations in F. For R of $10$, a change in the diffusion response becomes tangible with increasing F (from $0.2$ to $0.6$), characterized by a two-stage diffusion: an initial rapid diffusion stage followed by a marked deceleration. This indicates a deviation from Fickian behavior from a macroscopic perspective. However, with a further increase in F ($0.8$ to $1$), the diffusion curve realigns with the Fickian model. This two-stage diffusion phenomenon becomes even more pronounced at R of $60$ and $100$. For instance, at $R = 60$ with $F=0.2$, it illustrates a clear departure from Fickian diffusion, corroborated by the delineation of non-Fickian behavior in the boundary map shown in  Fig~\ref{fig:Boundary map D33 model 1-2-1} for these parameters. It is imperative to emphasize that the apparent Fickian and non-Fickian behaviors discussed herein emerge from the effect of multi-layers. Importantly, the observed non-Fickian behavior does not contradict the fundamental principles of Fickian diffusion, as each individual layer remains consistent with Fickian diffusion mechanisms.

\begin{figure}[h!]
    \centering
    \includegraphics[width=0.6\linewidth]{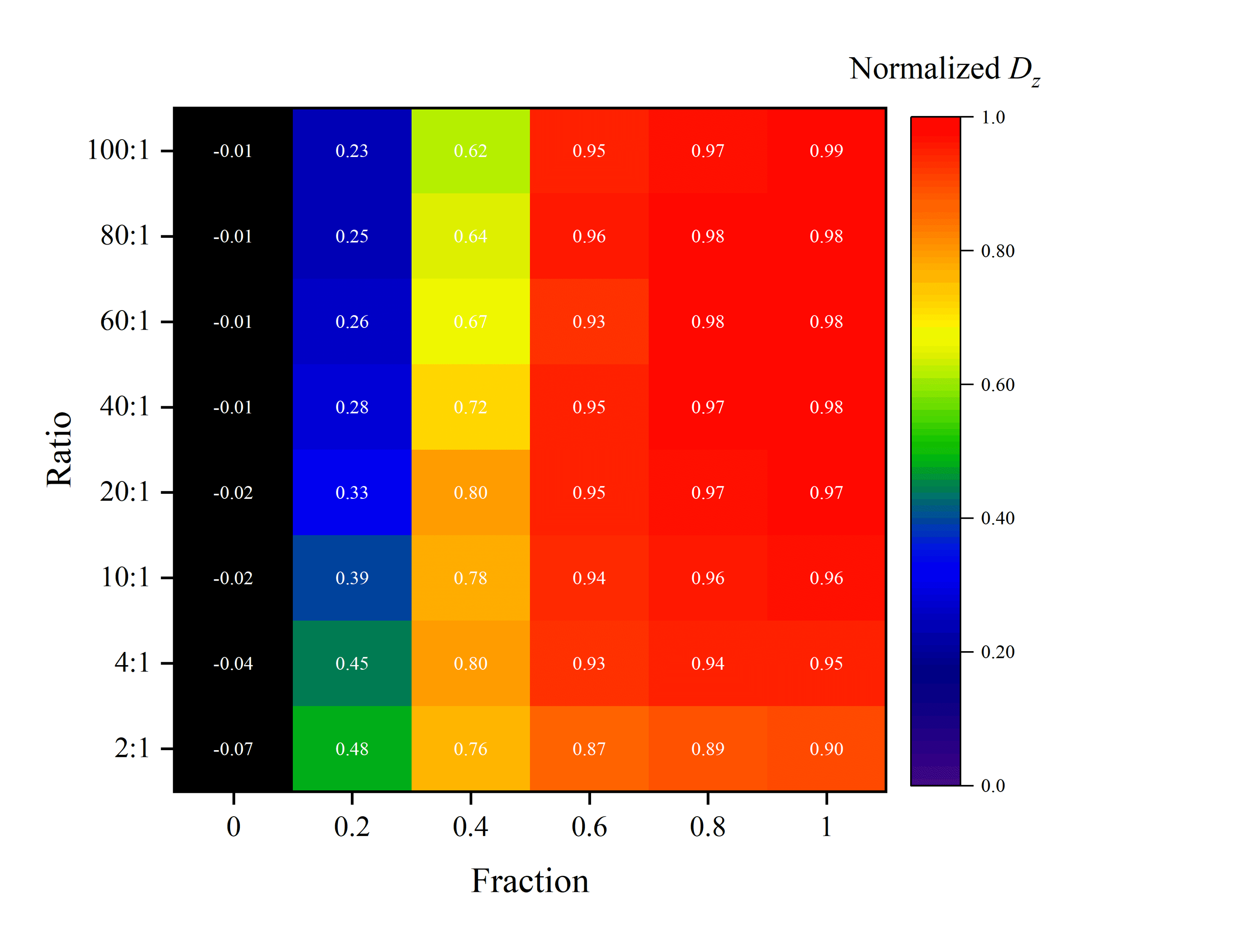}
    \caption{ Heat map of normalized $D_{z}$ with respect to R and F from Model 1-2-1}
    \label{fig:heat map D33 model 1-2-1}
\end{figure}

Fig~\ref{fig:heat map D33 model 1-2-1} shows the heat map of normalized $D_{z}$ with respect to F and R variables, illustrating the behavior of Model 1-2-1. It is observed that irrespective of variations in the R, the normalized $D_{z}$ dominantly approximates $0$ only when the F is lower. Conversely, in most scenarios, the normalized $D_{z}$ gravitates towards $1$. This indicates that within the 1-2-1 stacking order, where the diffusion coefficient of layer 1 is orders of magnitude higher than layer 2, the total diffusion coefficient of the multi-layered composite system is dominantly influenced by the outer (layer 1), which possesses the higher diffusion coefficient. 

\subsubsection{Model 2-1-2}

Linear regression analysis of the FEM simulation data for Model 2-1-2 demonstrates that the coefficient of determination ($R^2$) consistently exceeds $0.98$ as shown in Fig~\ref{fig:R2 model 2-1-2}. Within Model 2-1-2, the total diffusion behavior of the multi-layered composite system aligns with the one-dimensional Fickian diffusion model, irrespective of the variations in R and F. Consequently, the utilization of the one-dimensional Fickian diffusion approximation for calculating the total diffusion coefficients of multi-layered composite systems is validated by a high degree of accuracy.

\begin{figure}[h!]
    \centering
    \includegraphics[width=0.6\linewidth]{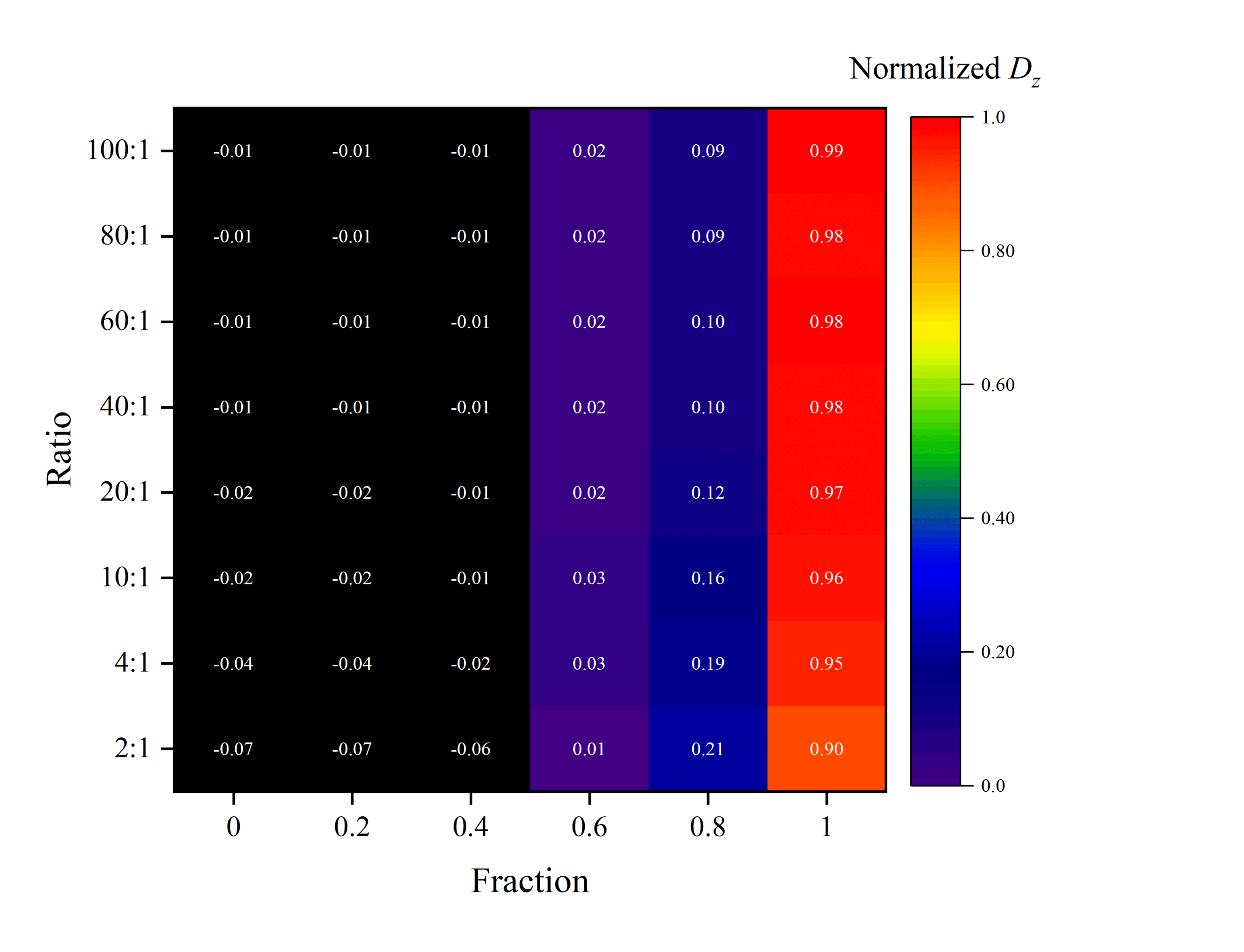}
    \caption{Heat map of normalized $D_{z}$ with respect to R and F for Model 2-1-2 in 33 direction}
    \label{fig:heat map D33 model 2-1-2}
\end{figure}

Fig~\ref{fig:FEM result D33 model 2-1-2} presents the results from simulations and curve fitting for Model 2-1-2 with R of $2, 10, 60, $and $100$. It can be seen that the multi-layered composites always exhibit Fickian behavior, regardless of the variations of R and F. A minor deviation is observed in the latter stages of diffusion, where simulation data slightly falls below the theoretical curve of the Fickian model. The difference could be attributed to the interior layer, layer 1, which possesses a higher diffusion coefficient and thus contributes more significantly during the final diffusion period. As seen in Fig~\ref{fig:heat map D33 model 2-1-2}, in the case of Model 2-1-2, the normalized $D_{z}$ is almost close to $0$ regardless of the variation of R at lower F values. Only when the F is high enough, the normalized $D_{z}$ is close to $1$. The apparent diffusion coefficient of Model 2-1-2 is also mainly determined by the outer layer (the layer with the smaller diffusion coefficient), which is consistent with the conclusion of Model 1-2-1. This conclusion can also explain how and why waterproofing, which has a lower diffusion coefficient and thinner thickness, can defend against moisture diffusion in FRPCs effectively.

\subsubsection{Comparison between Model 1-2-1 and Model 2-1-2}

\begin{figure}[h!] 
  \centering
  \subfigure[]
  {
      \label{fig:Comparison of curves}\includegraphics[width=0.48\linewidth]{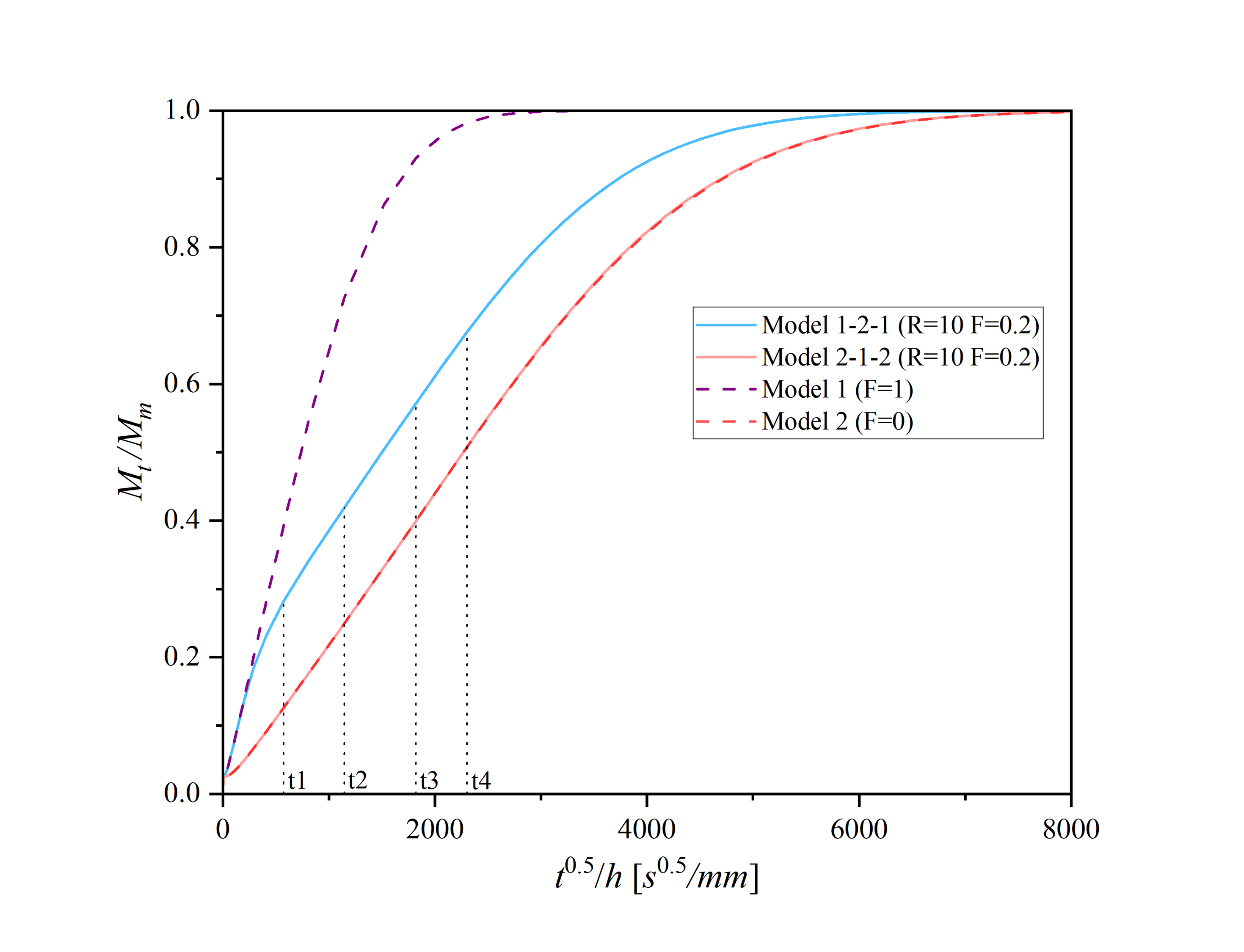}
  }
  \subfigure[]
  {
      \label{fig:Comparison of flux}\includegraphics[width=0.48\linewidth]{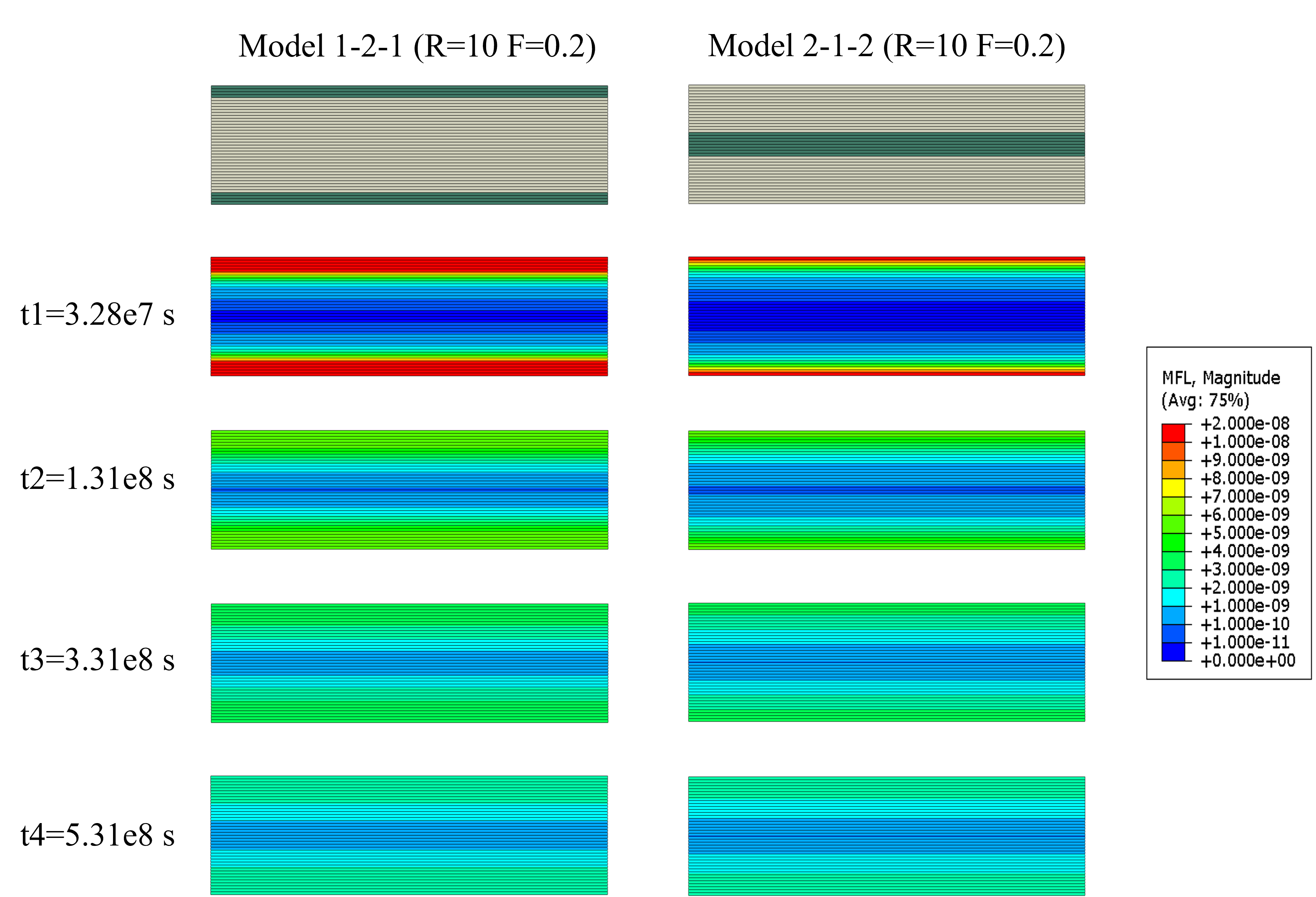}
  }
  \caption{ (a) Comparison of simulation data for different models. (b) Mass flux across the cross-section for Model 1-2-1 and Model 2-1-2 with moisture exposure time}
  \label{fig:Comparison of 4 FEM models} 
\end{figure}

From the FEM simulations, it appears that the diffusion coefficient of the outer layer dominantly influences the total diffusion coefficient of multi-layered composite systems. Results from 4 scenarios: 1) Model 1-2-1 (R=10, F=0.2), 2) Model 2-1-2 (R=10, F=0.2), 3) Model 1 (Material 1), and 4) Model 2 (Material 2) are examined to highlight the influence of outer layer on the total diffusion coefficient of multi-layered composite systems as shown in Fig~\ref{fig:Comparison of 4 FEM models}. It is worth reminding here that the diffusion coefficient of Material 1 is higher than that of Material 2.

In Fig~\ref{fig:Comparison of curves}, Model 1-2-1 exhibits a more rapid diffusion rate in the initial stages than Model 2-1-2, characterized by a two-stage diffusion behavior. During the initial stage, Model 1-2-1 has a significantly higher diffusion rate, closely matching that of Model 1. Conversely, the diffusion rate of Model 2-1-2 is lower, similar to the behavior of Model 2. Upon reaching a specific characteristic time, the diffusion rate of Model 1-2-1 rapidly decreases, and the slope of the curve becomes nearly identical to that of Model 2-1-2. During the entire diffusion process, Model 2-1-2 maintains a diffusion rate consistent with Model 2, with only minimal deviations during the latter diffusion stage.

The mass flux for both Model 1-2-1 (R=10, F=0.2) and Model 2-1-2 (R=10, F=0.2) at four distinct time points, $t_1, t_2, t_3,$ and $t_4$, across their cross-sections are illustrated in Fig~\ref{fig:Comparison of flux}. Initially, at $t_1$, Model 1-2-1 exhibits a significantly higher mass flux than Model 2-1-2, with the difference diminishing progressively at $t_2$ and $t_3$. By $t_4$, the mass fluxes of both models converge, indicating a comparable diffusion rate at this period. These images also corroborate the substantial impact of the outer layer on the early stage of the entire diffusion within multi-layered composite systems. Influenced by the concentration gradient of the diffusion medium, the rate of apparent diffusion is mainly determined by the initial stage. Therefore, {\bf the effect of stacking order on the total diffusion coefficient of multi-layered composite systems is extremely critical and cannot be neglected. As mentioned above, neither the series nor parallel models account for the stacking order.}

\begin{figure}[h!] 
  \centering
  \subfigure[]
  {
      \label{fig:D33 R=2 predicted Model}\includegraphics[width=0.48\linewidth]{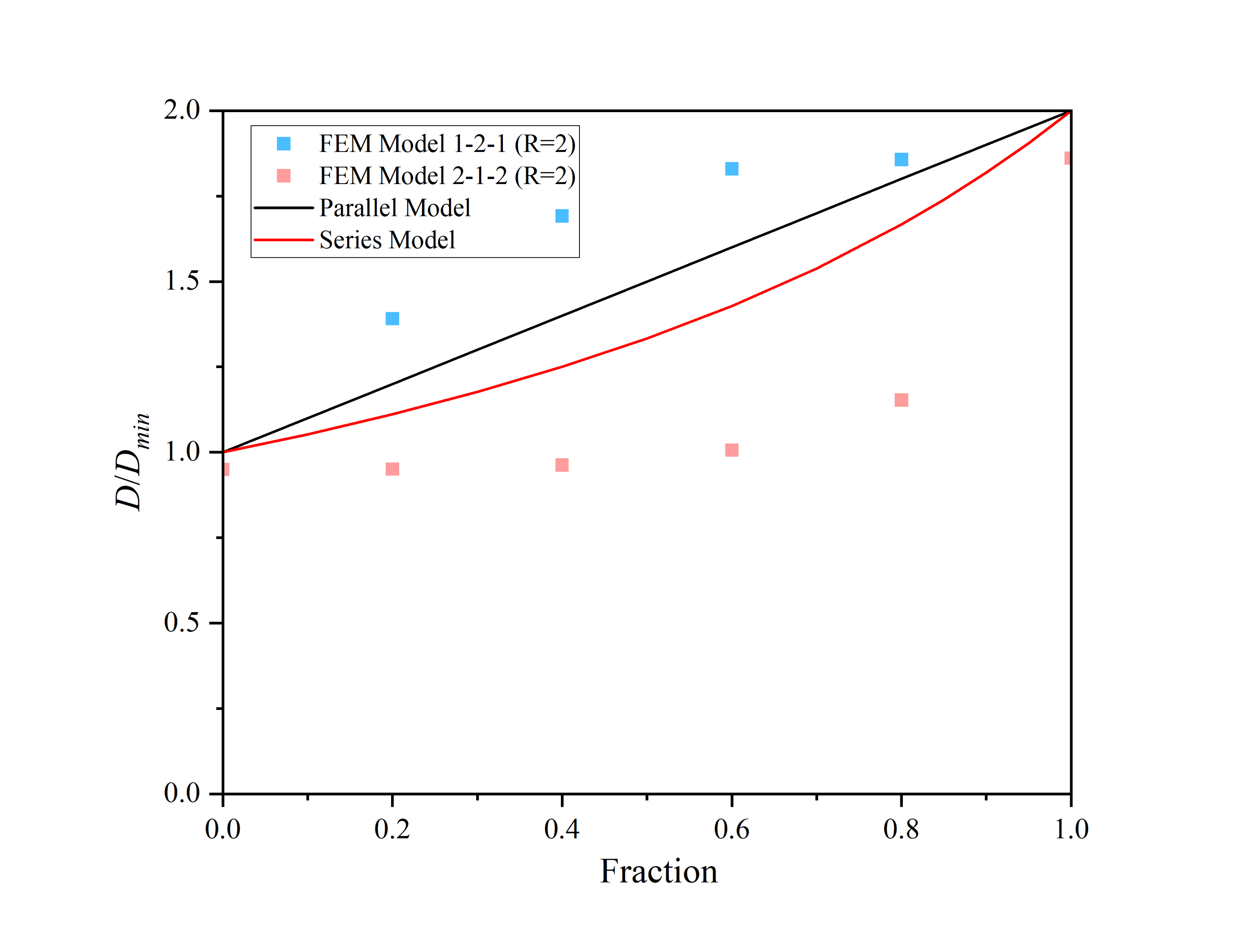}
  }
  \subfigure[]
  {
      \label{fig:D33 R=10 predicted Model}\includegraphics[width=0.48\linewidth]{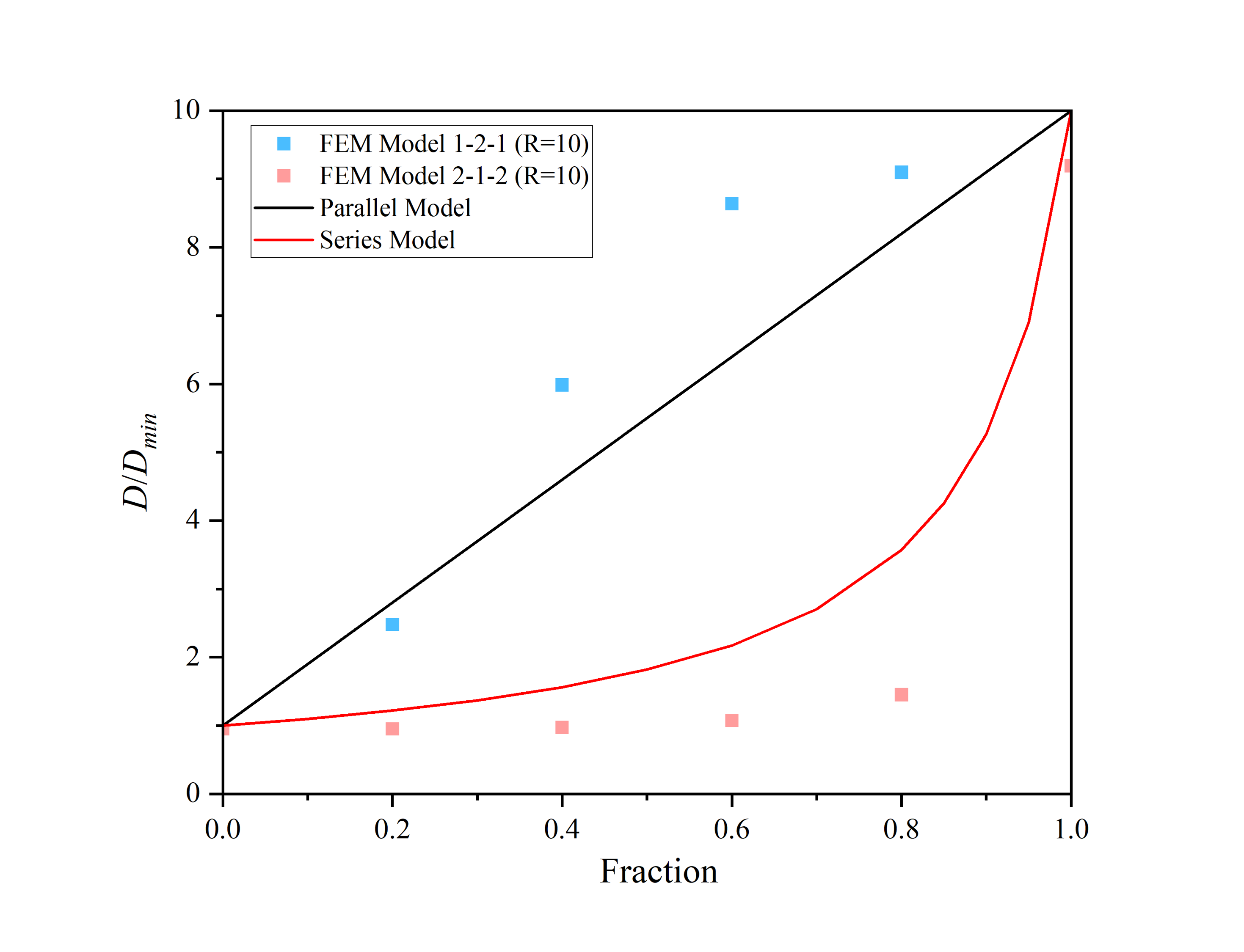}
  }
    \subfigure[]
  {
      \label{fig:D33 R=60 predicted Model}\includegraphics[width=0.48\linewidth]{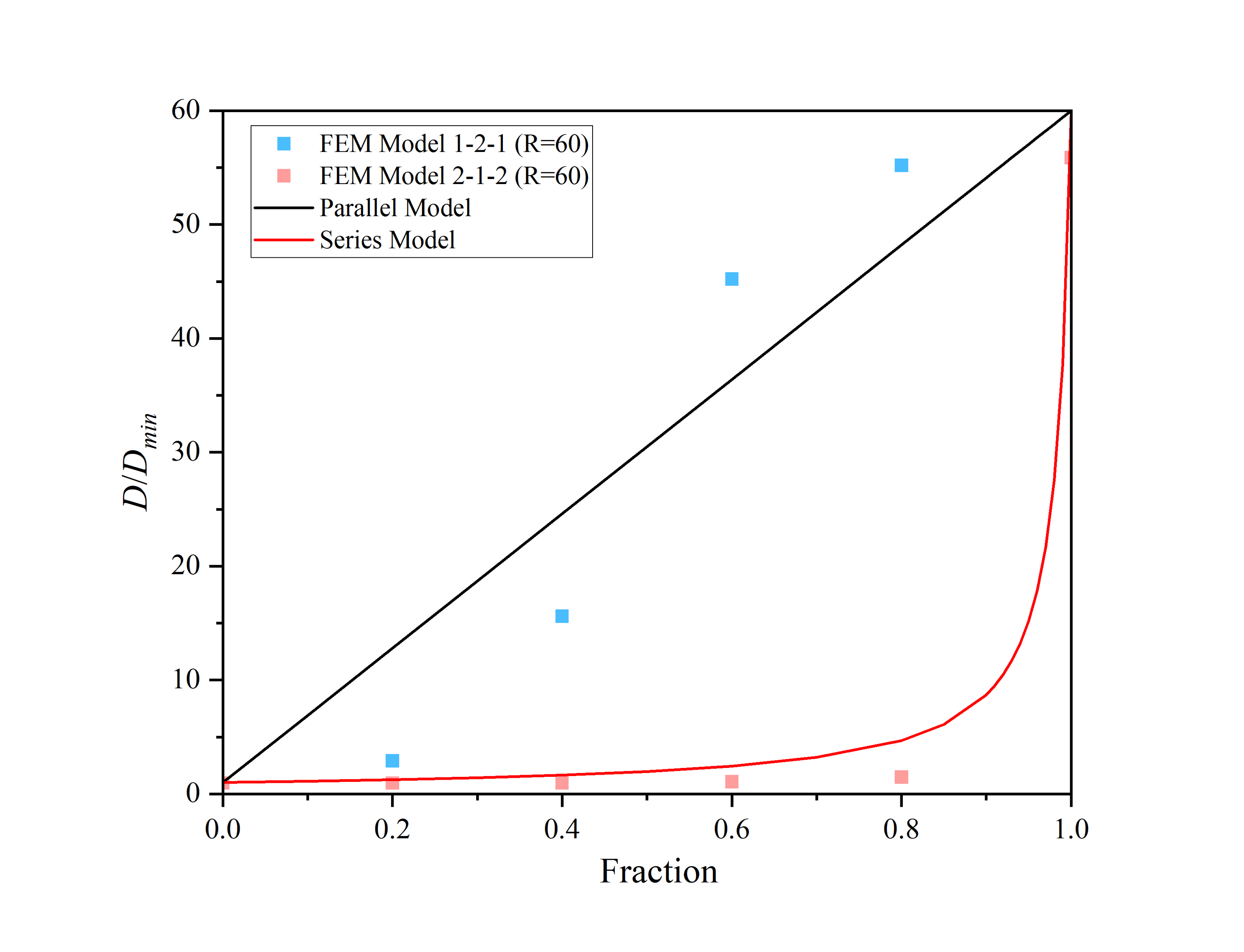}
  }
  \subfigure[]
  {
      \label{fig:D33 R=100 predicted Model}\includegraphics[width=0.48\linewidth]{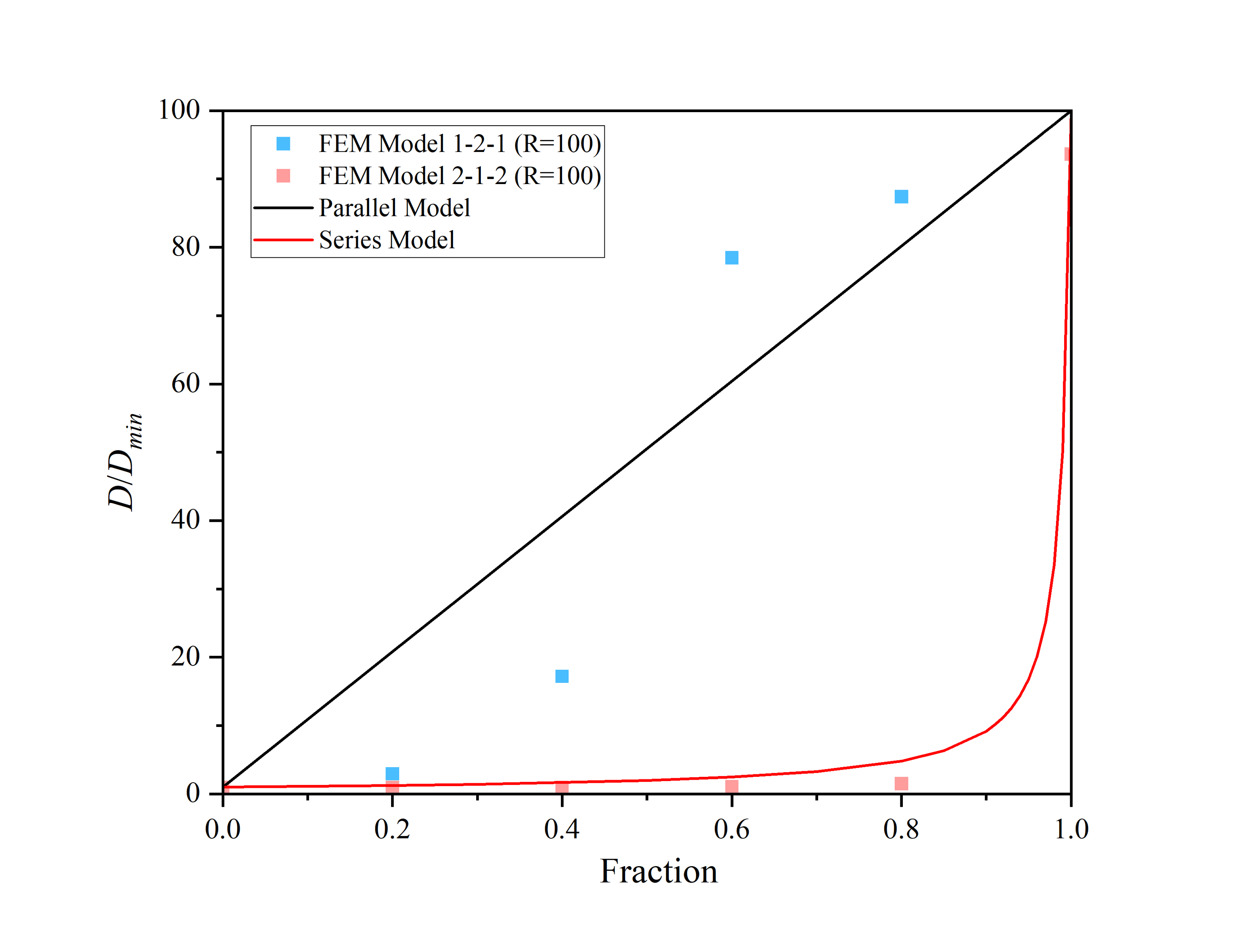}
  }
  \caption{ Calculated total diffusion coefficients of Model 1-2-1 and Model 2-1-2 in $z$ direction along with the parallel and series model (a) R=2 (b) R=10 (c) R=60 (D) R=100.}
  \label{fig:parallel and series model D33} 
\end{figure}

\subsubsection{New proposed model for $D_z$ of multi-layered composites}

Fig~\ref{fig:parallel and series model D33} presents the calculated total diffusion coefficients of Model 1-2-1 and Model 2-1-2 for R equal to $2, 10, 60$, and $100$, along with the theoretical predictions determined from the series and parallel models. It can be seen that the results of R=2 are not consistent with either the series or the parallel model, but in general, Model 1-2-1 is close to the parallel model, and Model 2-1-2 is close to the series model. With R increasing, Model 2-1-2 approaches the series model, but this may be due to the effect of the outer layer. The simulation results of Model 1-2-1 always fluctuate above and below the parallel model. Therefore, it is necessary to propose a new methodology to calculate the total diffusion coefficient of multi-layered composite systems accounting for layer stacking.

\begin{figure}[h!] 
  \centering
  \subfigure[]
  {
      \label{fig:D of model 1-2-1 and model 2-1-2}\includegraphics[width=0.48\linewidth]{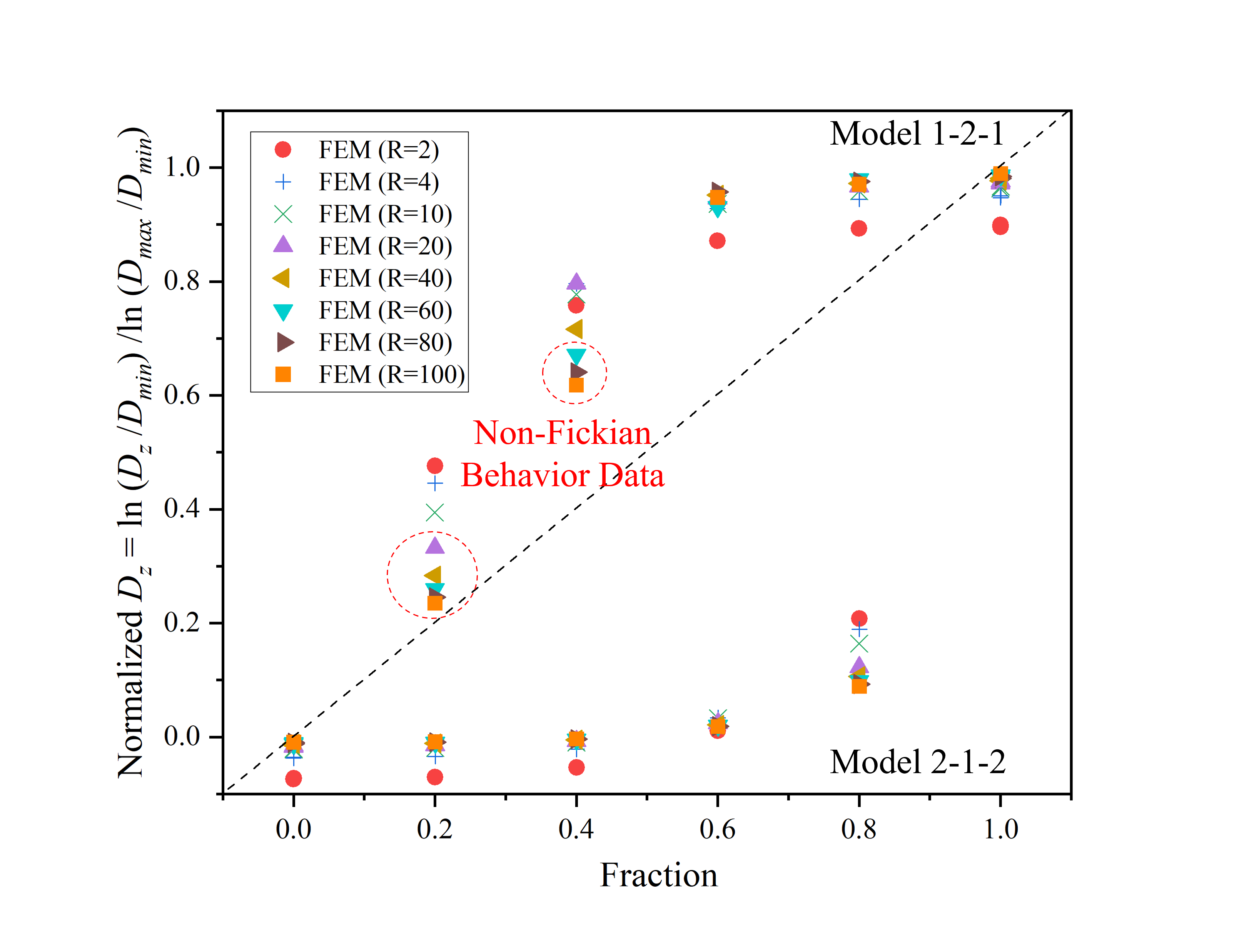}
  }
  \subfigure[]
  {
      \label{fig:new model}\includegraphics[width=0.48\linewidth]{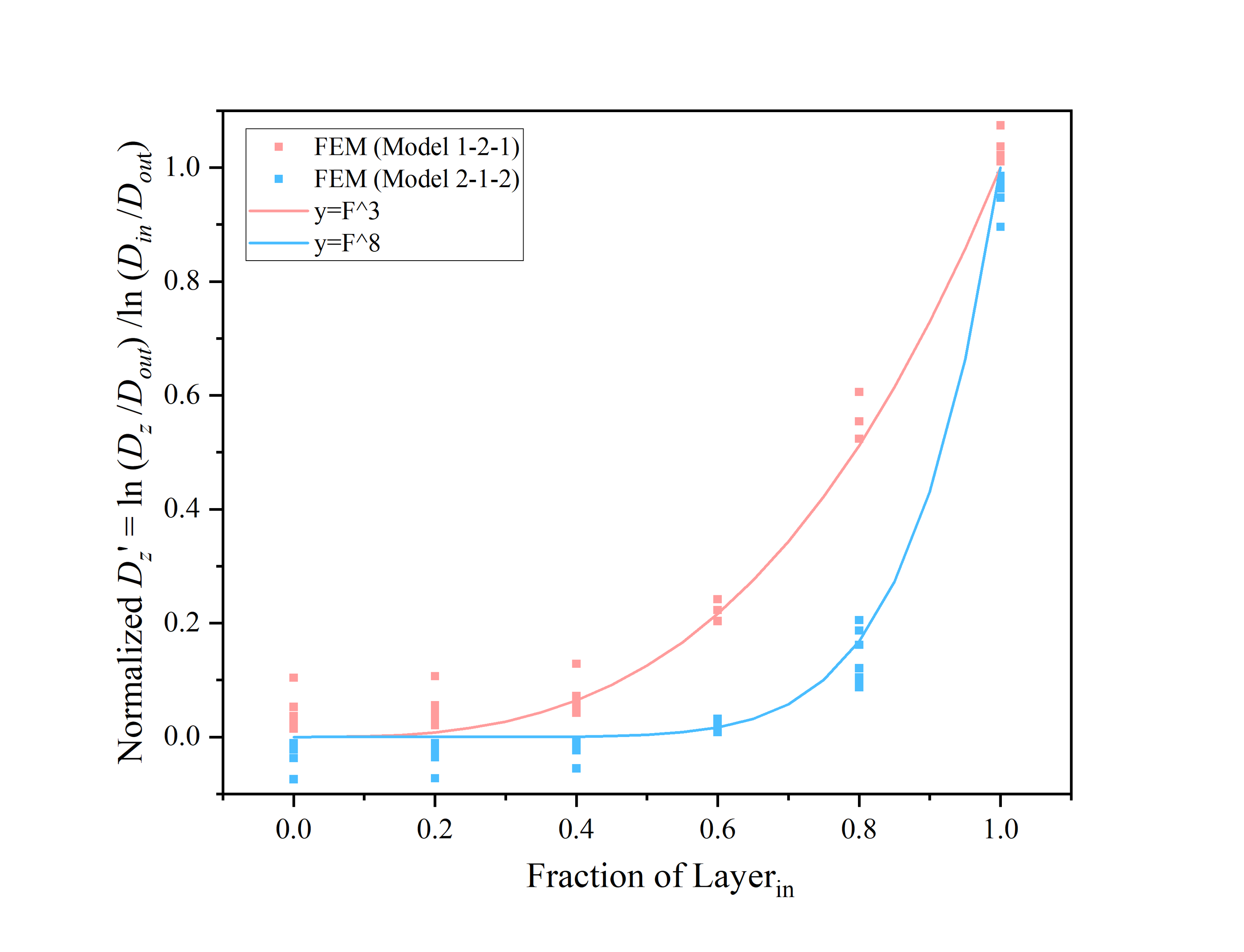}
  }
  \caption{ (a) Calculated normalized $D_{z}$ of Model 1-2-1 and Model 2-1-2 in the $z$ direction. (b) A new model is proposed to calculate the normalized $D'_{z}$ in the $z$ direction.}
  \label{fig:FEM data and new model} 
\end{figure}

Fig~\ref{fig:D of model 1-2-1 and model 2-1-2} shows the normalized $D_{z}$ (Eq \ref{eq:normalized D by min and max}) determined by the Fickian model for all simulation data of Model 1-2-1 and Model 2-1-2. The data points falling within the non-Fickian behavior are highlighted in the plot. 

To consider the effect of stacking order, the diffusion coefficient is normalized according to a {\bf new proposed expression} shown in Eq~\ref{eq:normalized D by in and out},

\begin{equation}\label{eq:normalized D by in and out}    
  Normalized \: D^{'}= \frac{\ln{⁡(D/D_{out})}}{\ln{⁡(D_{in}/D_{out})}}
\end{equation}

where $D_{in}$ and $D_{out}$ are the diffusion coefficients of the inner and outer layers, respectively, in the multi-layered materials instead of considering the min and max values of $D$. 

The plot of normalized $D'_{z}$ after omitting the data from the non-Fickian behavior region is shown in Fig~\ref{fig:new model}. It is observed that the data of Models 1-2-1 and 2-1-2 do not overlap, but display analogous trends. A power function is used to fit the data as shown in Eq~\ref{eq:new model}, 

\begin{equation}\label{eq:new model}
\frac{\ln{⁡(D/D_{out})}}{\ln{⁡(D_{in}/D_{out})}}=F^n
\end{equation}

where $n$ equals $3$ and $8$ for Model 1-2-1 and Model 2-1-2, respectively. $F$ is the fraction of the inner layer. Model 1-2-1 adheres to this power function model within the domain of $R^2 > 0.9$, which means that the multi-layered composite system should conform to quasi-Fickian or Fickian behavior. Model 2-1-2 has shown to inherently follow a Fickian behavior regardless of the values of R and F. More accurate than the series and parallel models, the power function model comprehensively accounts for the influence of stacking order on the total diffusion coefficient. In Eq~\ref{eq:new model}, $F$ and $n$ symbolize the influence of the layer volume fraction and stacking order, respectively.

\subsection{Experiments on Multi-Layered Pultruded Plate}

\begin{figure}[h!]
    \centering
    \includegraphics[width=0.6\textwidth]{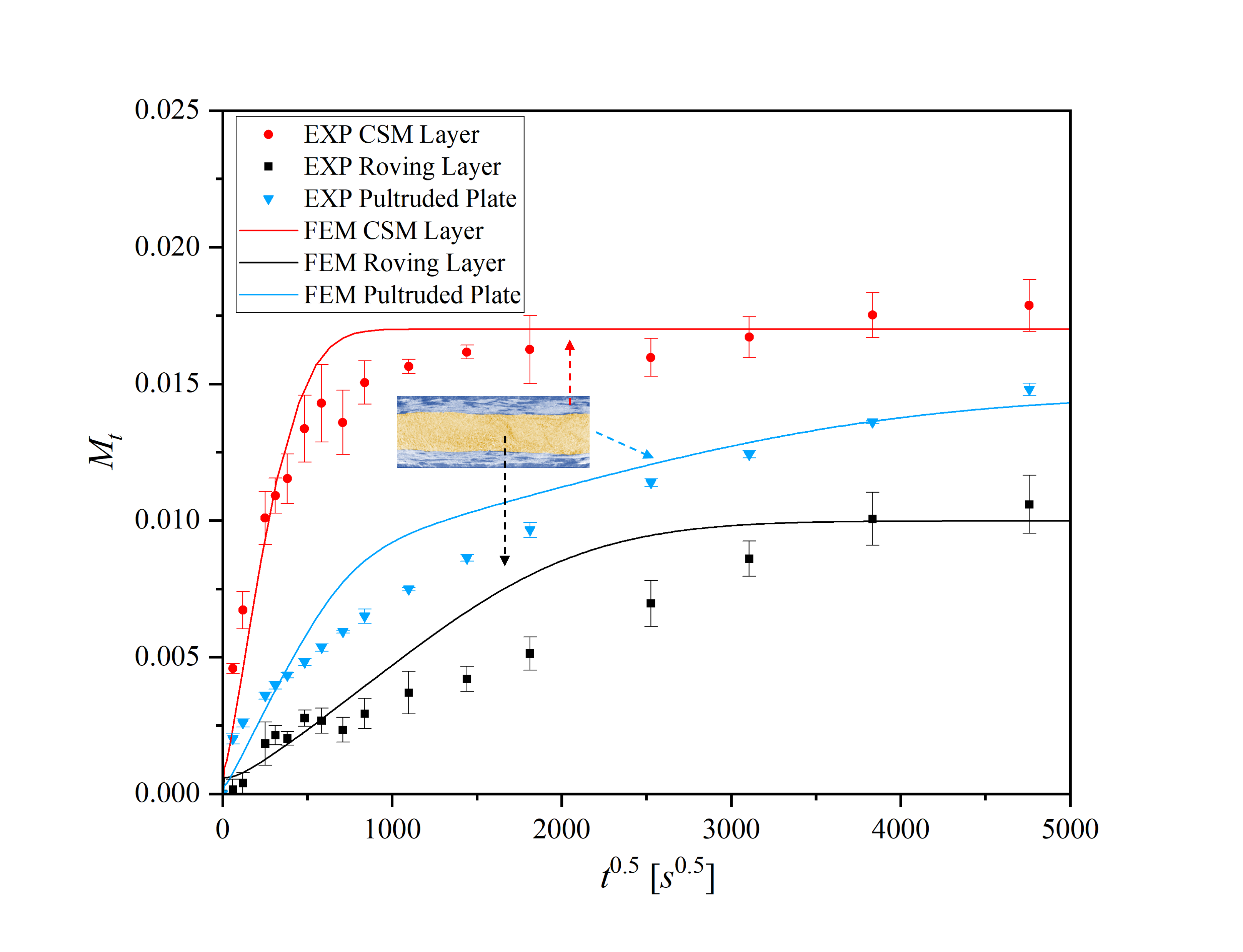}
    \caption{Experimental and FEM simulation results of Multi-layered pultruded plate (Model 1-2-1).}
    \label{fig:result of Multi-layered pultruded plate}
\end{figure}

Fig~\ref{fig:result of Multi-layered pultruded plate} shows the weight gain data of a multi-layered pultruded plate under water immersion test at room temperature. The outer CSM layer, which reaches saturation at $0.017$, is considered to follow the Fickian Diffusion Model. However, it appears that the pultruded plate has not yet reached saturation, and we assume it will be saturated at $0.014$. The pultruded plate has an apparent two-stage diffusion behavior related to the significant difference in diffusion coefficients of the two material layers, consistent with the conclusion above for Model 1-2-1. Parameters calculated from experiments are given as Table \ref{tab:parameter of pultruded for fem} and are also used in the FEM simulation. 

\begin{table}[h!]
    \centering
        \caption{Parameters of the multi-layered pultruded plate and corresponding individual materials}
    \begin{threeparttable}
    \begin{tabular}{|c|c|c|c|c|} \hline  
         Specimen&  $\rho$ $(kg/m^3)$ &  $M_m$ &  $D$ $(mm^2/s)$&  $s$ $(ppm)$  \\ \hline  
         CSM Layer&  $1699.92$& $0.017$& $9.52\times 10^{-7}$& $2.89\times 10^4$\\ \hline  
         Roving Layer& $1832.25$& $0.009$& $3.36\times 10^{-8}$& $1.65\times 10^4$\\ \hline  
         Pultruded Plate& -& $0.014$\tnote{1}& -& -\\\hline
    \end{tabular}
    \begin{tablenotes}
    \footnotesize
    \item[1] Assumed from current experiment data.
    \end{tablenotes}
    \label{tab:parameter of pultruded for fem}
    \end{threeparttable}
\end{table}

Simulation results of the CSM and Roving layers follow the Fickian diffusion model, which coincides with their experiment results. The simulated weight gain curve of the pultruded plate exhibits two-stage diffusion behavior, slightly delayed behind the experiment result. The thickness of the outer layer in the pultruded plate is only $0.74$ $mm$, while the thickness of the extracted CSM layer is $0.95$ $mm$. This could imply that the material of the CSM layer contains not only pure CSM but also the roving layer and their interface due to the precision of the CNC process. In this case, the diffusion coefficient of the CSM layer calculated from the experimental result is smaller than the actual one. For the pultruded plate, the actual diffusion coefficient of an outer layer in the experiment is higher than the calculated diffusion coefficient in the simulation. Hence, this could have contributed to an earlier onset of two-stage behavior in the experiment than in the simulation. Overall, experiment and simulation results demonstrate the two-stage behavior of Model 1-2-1, verifying the accuracy of the parametric analysis method.

\subsection{Effect of Solubility}\label{effect_of_solubility}
\begin{figure}[h!] 
  \centering
  \subfigure[]
  {
      \label{fig:solubility Model 1-2-1}\includegraphics[width=0.48\linewidth]{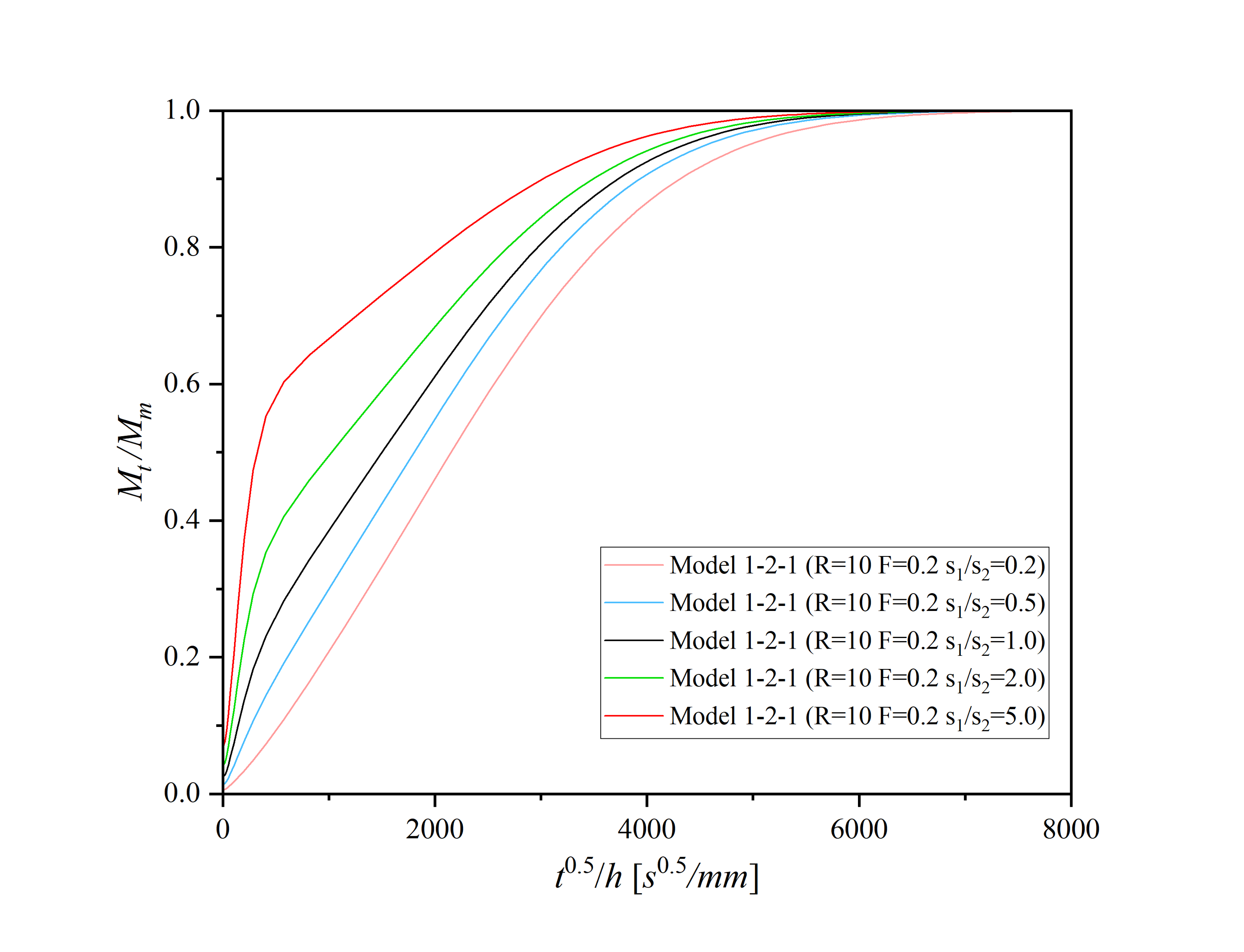}
  }
  \subfigure[]
  {
      \label{fig:solubility Model 2-1-2}\includegraphics[width=0.48\linewidth]{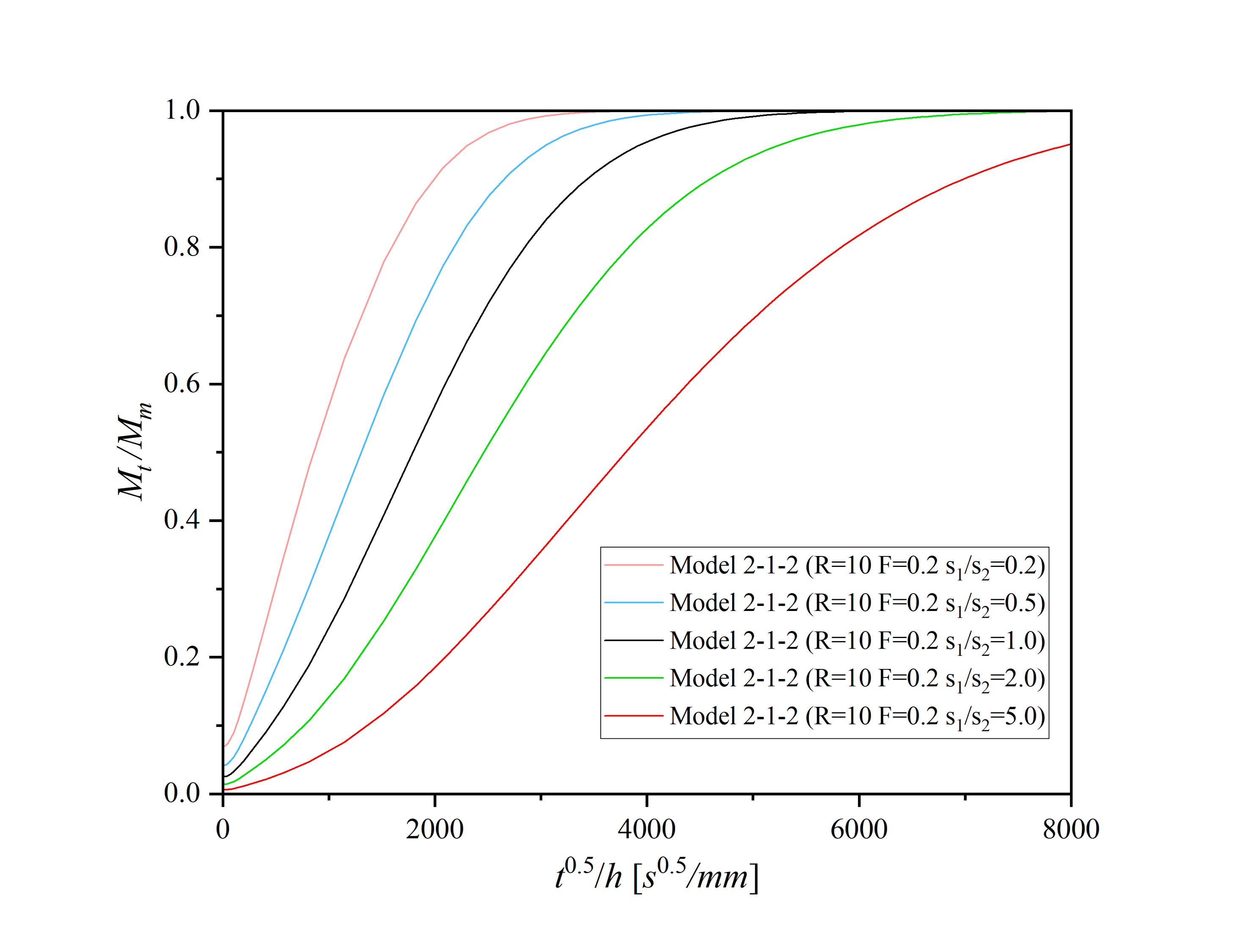}
  }
  \caption{ Simulation results of weight gain with different $s_1 / s_2 $ in the $z$ direction, where $s_2$ set to $1$: (a) Model 1-2-1 (b) Model 2-1-2.}
  \label{fig:solubility comparsion} 
\end{figure}

In practice, the solubility of individual materials in a multi-layered system may not be the same as shown in Table\ref{tab:parameter of pultruded for fem}. Simulations of Model 1-2-1 and Model 2-1-2 with different $s_1 / s_2$ are conducted to discover the influence of solubility. Fig~\ref{fig:solubility Model 1-2-1} shows a two-stage diffusion behavior for Model 1-2-1 when $s_1 / s_2 = 1$. This two-stage behavior becomes more apparent with increasing $s_1 / s_2 $. However, the two-stage behavior disappears and follows a Fickian behavior when $s_1 / s_2 $ decreases below 1. From Fig~\ref{fig:solubility Model 2-1-2}, Model 2-1-2 always manifests Fickian behavior regardless of how $s_1 / s_2 $ changes. Nonetheless, the diffusion rate is slower as $s_1 / s_2 $ increases from $0.2$ to $5$.

The solubility of individual materials is generally different in practical multi-layered materials. The outer layer with higher solubility can advance the entire diffusion process faster. In particular, multi-layered materials like Model 1-2-1, whose outer layer has higher solubility and a higher diffusion coefficient, display a more apparent two-stage diffusion behavior. However, for most types of FRPCs, the solubility is not significantly different, which is mainly determined by the resin type and its volume fraction. So, the new model proposed above can be applied to most application scenarios of multi-layered FRPCs. 

\subsection{Design Methodology}
\begin{figure}[h!] 
  \centering
      \includegraphics[width=0.8\linewidth]{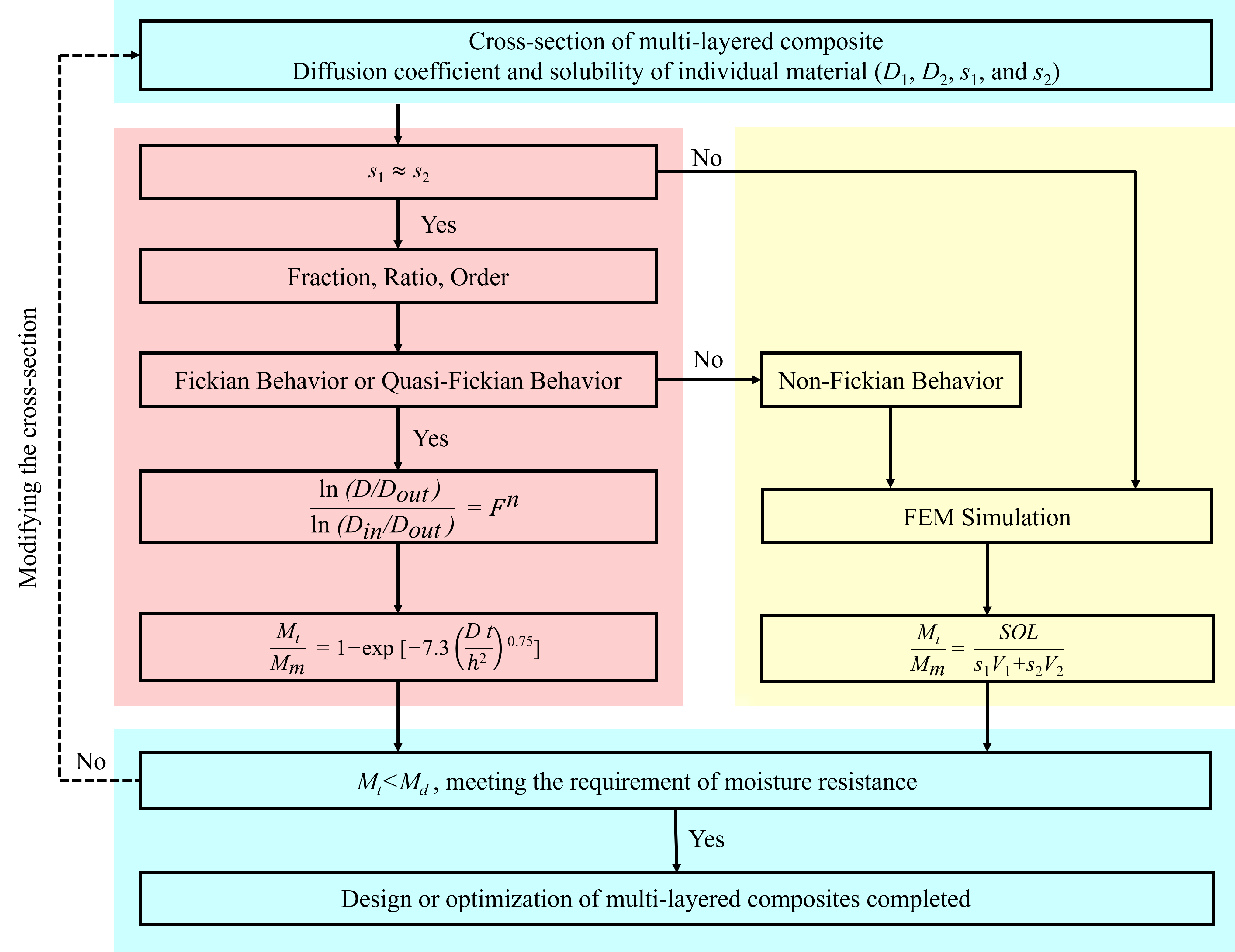}

  \caption{A methodology for designing and optimizing the cross-section of multi-layered materials considering moisture resistance.}
  \label{fig:design method} 
\end{figure}

Based on the simulation and experimental results, the methodology for designing and optimizing the cross-section of multi-layered materials considering long-time moisture resistance is presented in Fig~\ref{fig:design method}. For multi-layered materials with two types of materials, the diffusion behavior can be determined by the Fraction, Ratio, and Stacking Order if $s_1$ approximately equals $s_2$ according to Fig~\ref{fig:Boundary map D33 model 1-2-1} and Fig~\ref{fig:R2 model 2-1-2}(b). If it is a Fickian or Quasi-Fickian behavior, the total diffusion coefficient can be calculated based on the new model proposed according to Eq~\ref{eq:new model}. The weight gain can then be determined by the approximate curve of the one-dimensional Fickian model according to Eq~\ref{eq:approximation model}. In the case of non-Fickian behavior, a FEM simulation is essential to determine the overall moisture gain of the multi-layered material, which can also apply to cases where $s_1$ differs significantly from $s_2$. The cross-section can meet the requirements if the predicted moisture content at a specified time is less than the design value ($M_t < M_d$). Otherwise, the cross-section should be adjusted to meet the moisture content requirement. Overall, this design and optimization method, combined with the new proposed model and FEM simulation, is more efficient than only FEM simulation when considering the moisture resistance of multi-layered composites.

\section{Conclusions}\label{conc}

In the paper, we proposed three essential parameters that influence the total diffusion coefficient of multi-layered materials, especially for multi-layered FRPCs: the ratio of diffusion coefficients (Ratio or R), the volume fraction (Fraction or F), and the stacking order of individual layers. Through FEM simulations, we show that the stacking order can significantly influence the total diffusion coefficient of multi-layered materials, which both parallel and series theoretical models do not consider.

We conducted a parametric study using FEM simulations to determine how these three parameters influence the diffusion behavior of multi-layered materials. We chose two types of layer stacking: Model 1-2-1, with outer layers having higher diffusivity than the inner middle layer. The other is Model 2-1-2, which is the opposite of Model 1-2-1. Models 1-2-1 and 2-1-2 show Fickian diffusion behavior in the in-plane directions ($x$ or $y$), and the parallel model can calculate their total diffusion coefficients. In the $z$ direction, Model 1-2-1 manifests Fickian, quasi-Fickian, and non-Fickian diffusion behaviors based on the values of Ratio and Fraction. Conversely, regardless of how the Ratio and Fraction change, Model 2-1-2 consistently exhibits Fickian behavior. Neither the parallel nor the series model can determine the total diffusion coefficient in the $z$ direction of these multi-layered materials. So, we proposed a new model that accounts for layer stacking, which agrees well with the simulation results.

By comparing the weight gain curves of Model 1-2-1 with Model 2-1-2 at the same R and F, we found that even a very thin outer layer, with a much lower diffusion coefficient than the inner layer, can slow down the total diffusion process of multi-layered materials. This can also explain why waterproof coatings can effectively defend against moisture ingress towards composites. 

We performed a distilled water immersion test of a multi-layered pultruded plate at room temperature. The experimental results are in good agreement with the corresponding FEM simulation, demonstrating the accuracy of the parametric study. It should be noted that the solubility of individual materials differs from each other in actual multi-layered composites, which will also affect the total diffusion behavior. 

In summary, considering long-time moisture resistance, we proposed a methodology for designing and optimizing the cross-section of multi-layered materials, especially for multi-layered FRPCs. For multi-layered composite models with Fickian behavior, we can calculate the total diffusion coefficient directly using the newly proposed model than performing FEM simulation.

The multi-layered composite model considered in this paper is limited to three layers with symmetrical features. Solubility is not considered in particular. Future studies should consider multi-layered materials with more complicated stacking orders and the effect of solubility on the total diffusion coefficient. 

\section*{Acknowledgements}
The authors would like to acknowledge the support of the National Natural Science Foundation of China (No.U2106219) and Tsinghua Visiting Doctoral Students Foundation for supporting the research associate position for Shaojie Zhang to visit the University of Wisconsin-Madison to perform this study.

The authors would like to acknowledge the  support through the NSF CAREER award [\#: 2046476] through {\em{Mechanics of Materials and Structures (MOMS) Program}} for conducting the research presented here. 

\section*{Author Contributions}
S.Z. contributed to the conceptualization of the methodology, formal analysis, investigating, visualization, verification, and writing -- preparing the original draft. Y.L. contributed to the formal analysis and investigating. P.F. contributed to the supervision, writing -- reviewing and editing, and funding acquisition. P.P contributed to the conceptualization of the methodology, writing -- reviewing and editing, visualization, verification, supervision, project administration, and funding acquisition.

\section*{Data Availability}
The data that support the findings of this study are available from the corresponding author, PP, upon reasonable request.



{\footnotesize
\bibliographystyle{unsrt}
\bibliography{paper_bib}
}

\newpage
\appendix
\renewcommand\thefigure{\thesection.\arabic{figure}}
\setcounter{figure}{0}    

\section{Simulation of Moisture Diffusion in the $x$ direction ($D_{x}$)}

\begin{figure}[H] 
  \centering
  \subfigure[]
  {
      \label{fig:D11 model 1-2-1 R=2}\includegraphics[width=0.48\linewidth]{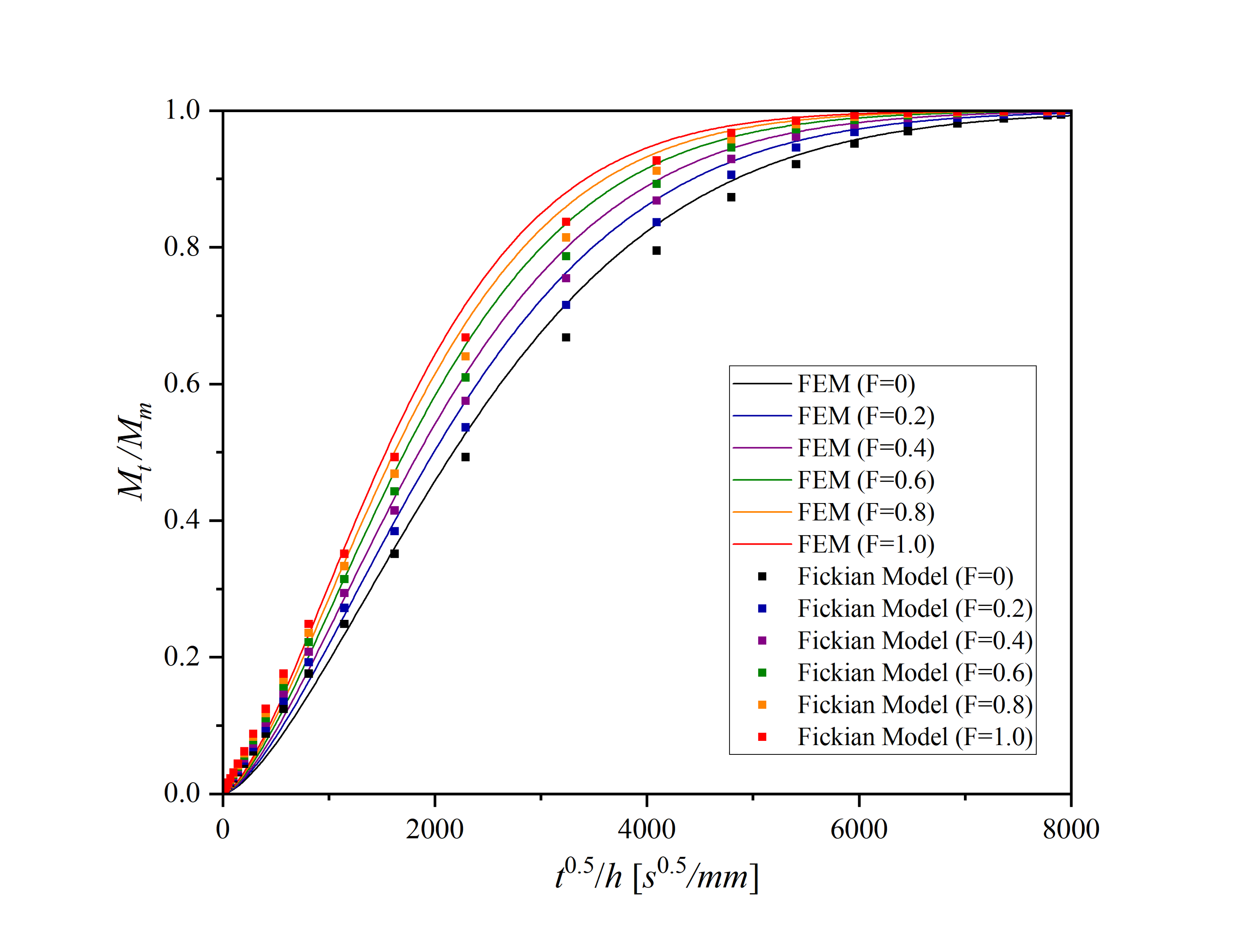}
  }
  \subfigure[]
  {
      \label{fig:D11 model 1-2-1 R=10}\includegraphics[width=0.48\linewidth]{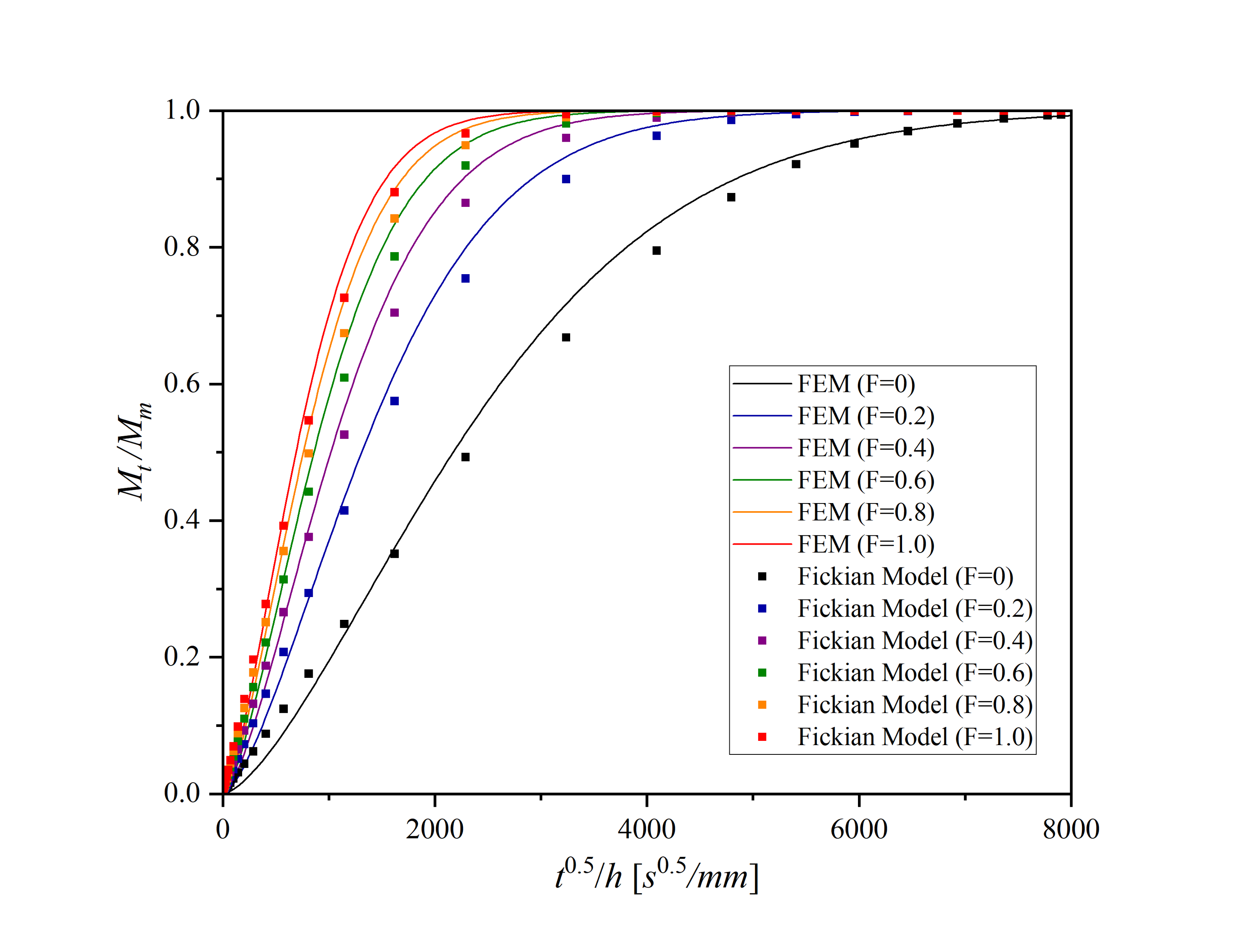}
  }
    \subfigure[]
  {
      \label{fig:D11 model 1-2-1 R=60}\includegraphics[width=0.48\linewidth]{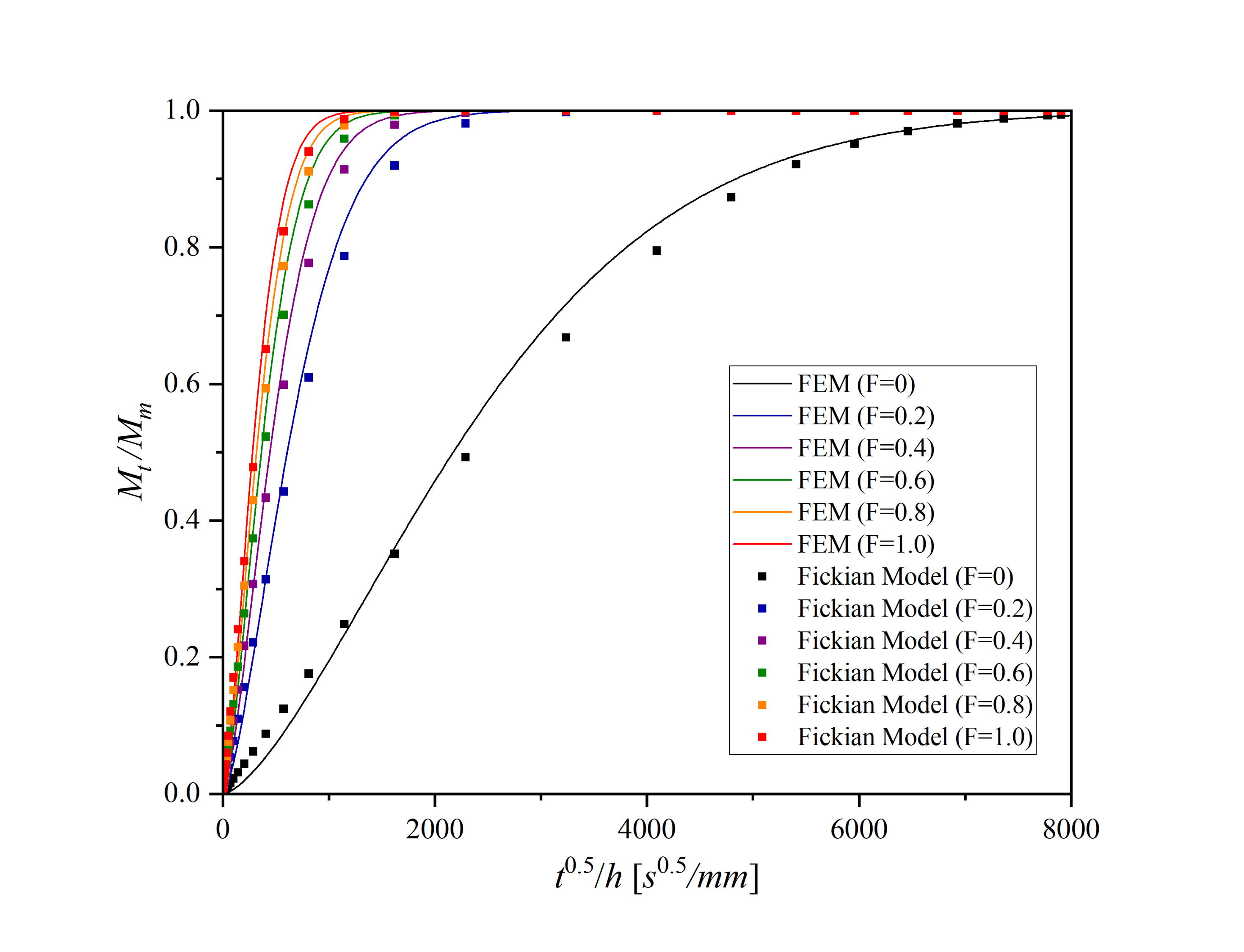}
  }
  \subfigure[]
  {
      \label{fig:D11 model 1-2-1 R=100}\includegraphics[width=0.48\linewidth]{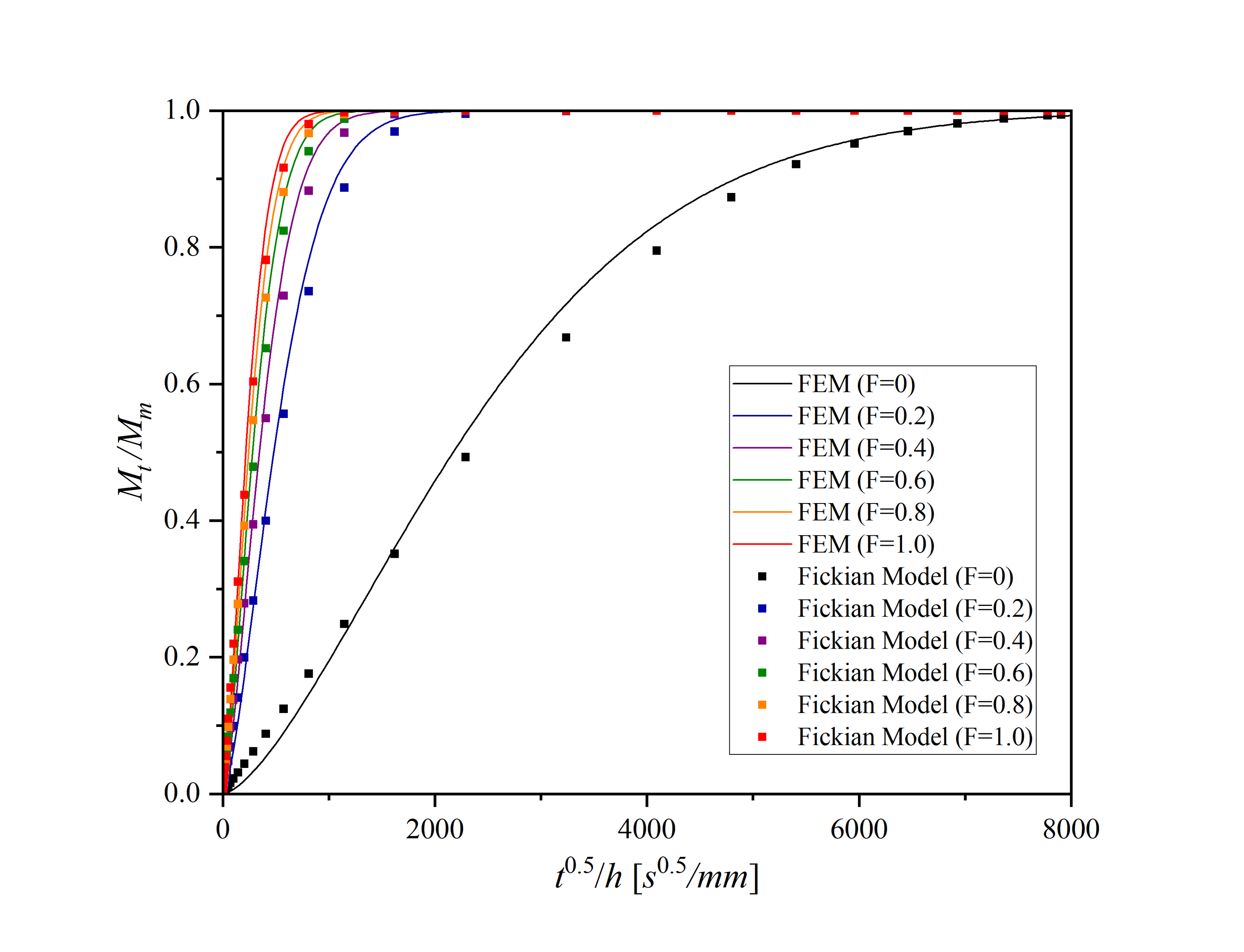}
  }
  \caption{ Results of simulation data and Fickian model from Model 1-2-1 in the $x$ direction (a) R=2 (b) R=10 (c) R=60 (d) R=100.}
  \label{fig:FEM result D11 model 1-2-1} 
\end{figure}

Fig~\ref{fig:FEM result D11 model 1-2-1} presents the result of Model 1-2-1 for R of $2$, $10$, $60$, and $100$, wherein the scatter plots represent the FEM simulation data. Continuous lines delineate the predictions based on the analytical Fickian model given in Eq~\ref{eq:linear model} and Eq~\ref{eq:approximation model}.

\begin{figure}[h!] 
  \centering
  \subfigure[]
  {
      \label{fig:D11 with parallel series model R=2}\includegraphics[width=0.48\linewidth]{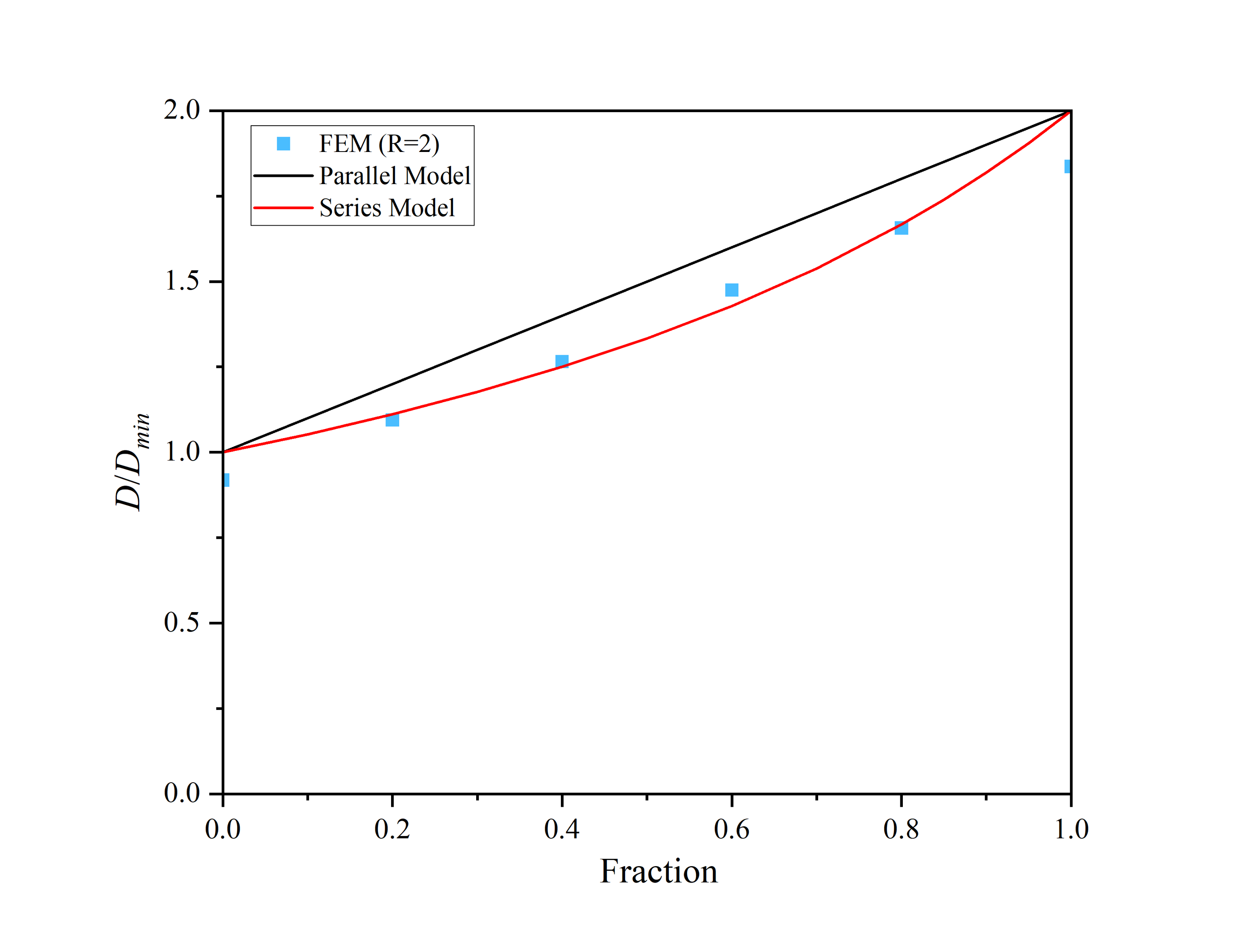}
  }
  \subfigure[]
  {
      \label{fig:D11 with parallel series model R=10}\includegraphics[width=0.48\linewidth]{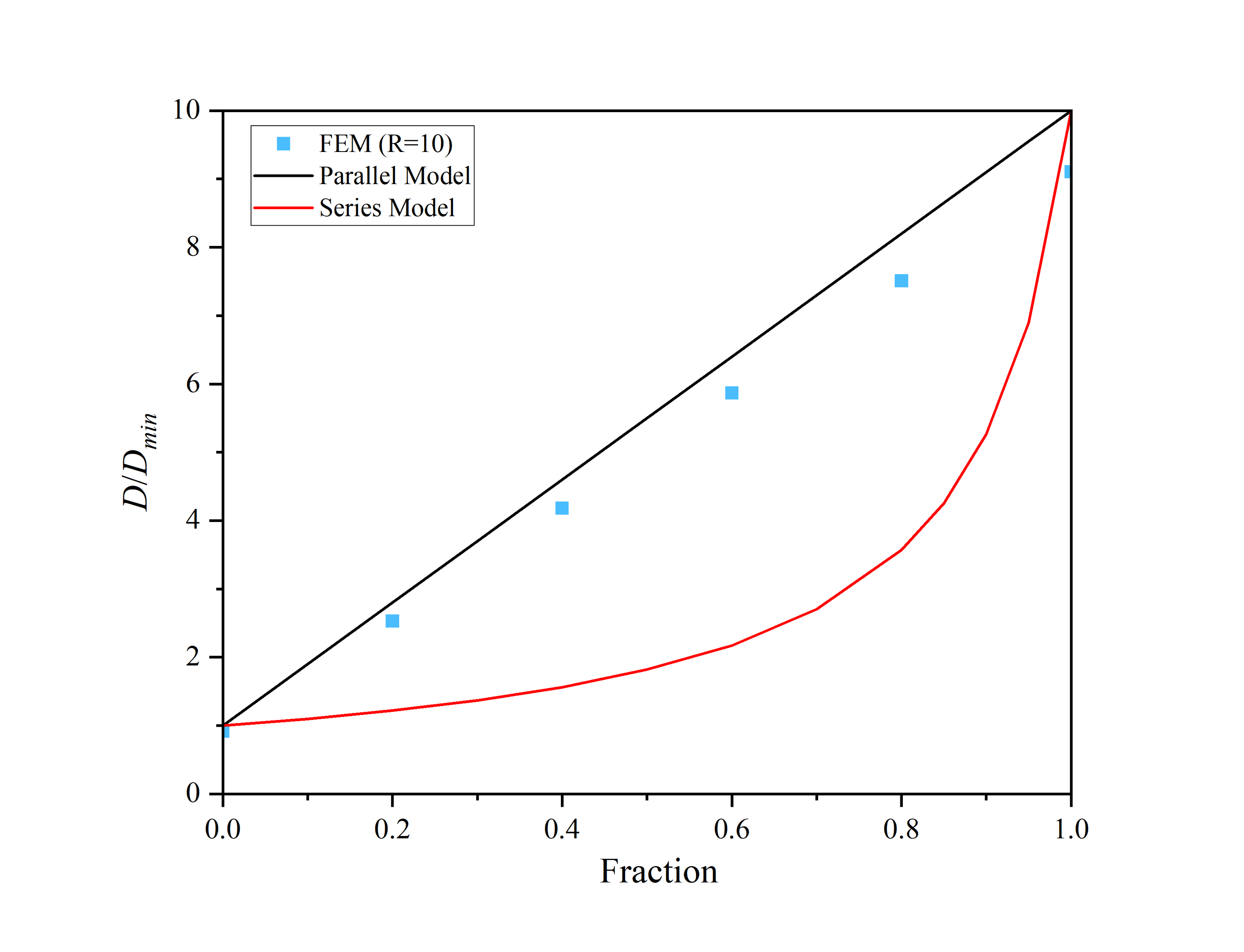}
  }
    \subfigure[]
  {
      \label{fig:D11 with parallel series model R=60}\includegraphics[width=0.48\linewidth]{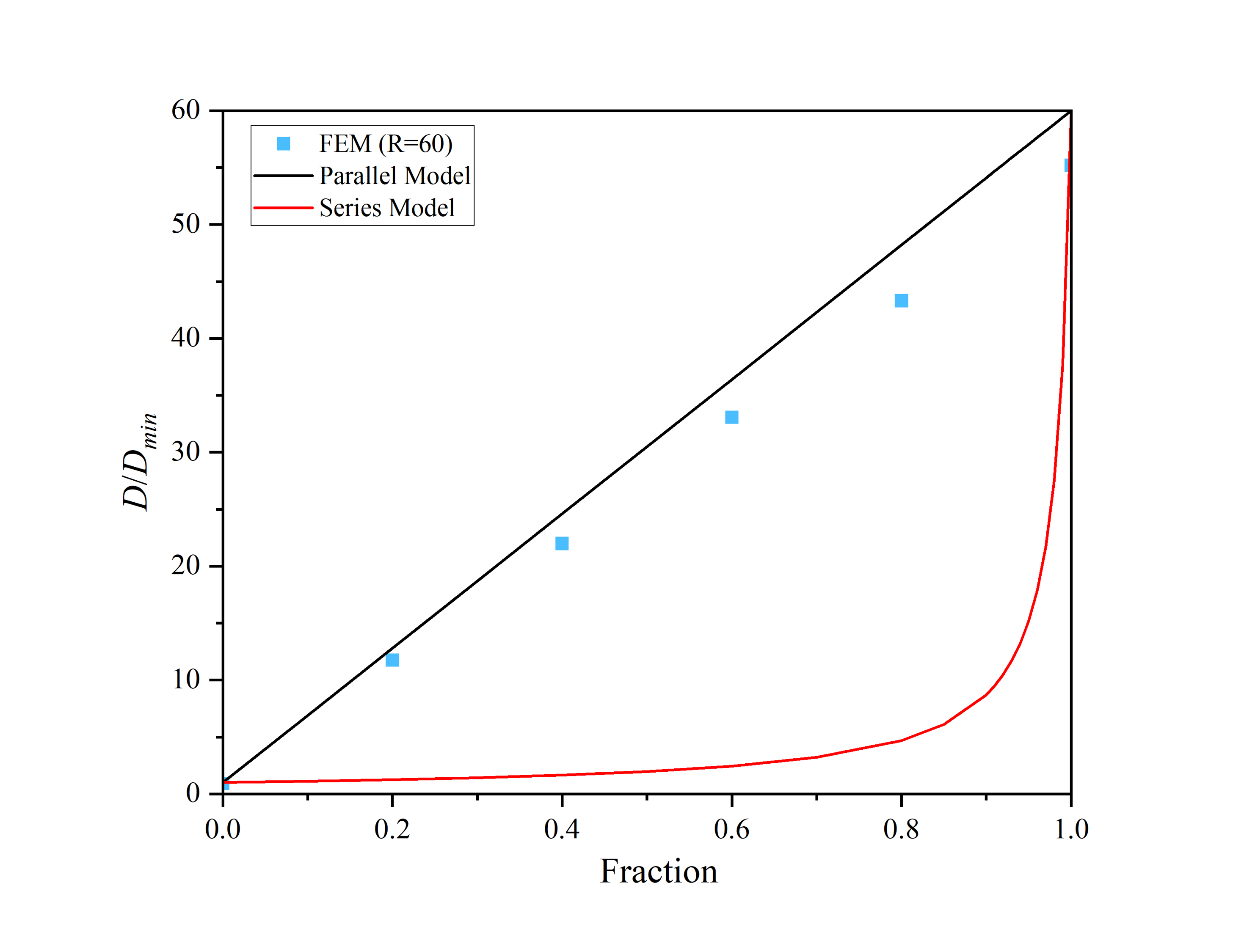}
  }
  \subfigure[]
  {
      \label{fig:D11 with parallel series model R=100}\includegraphics[width=0.48\linewidth]{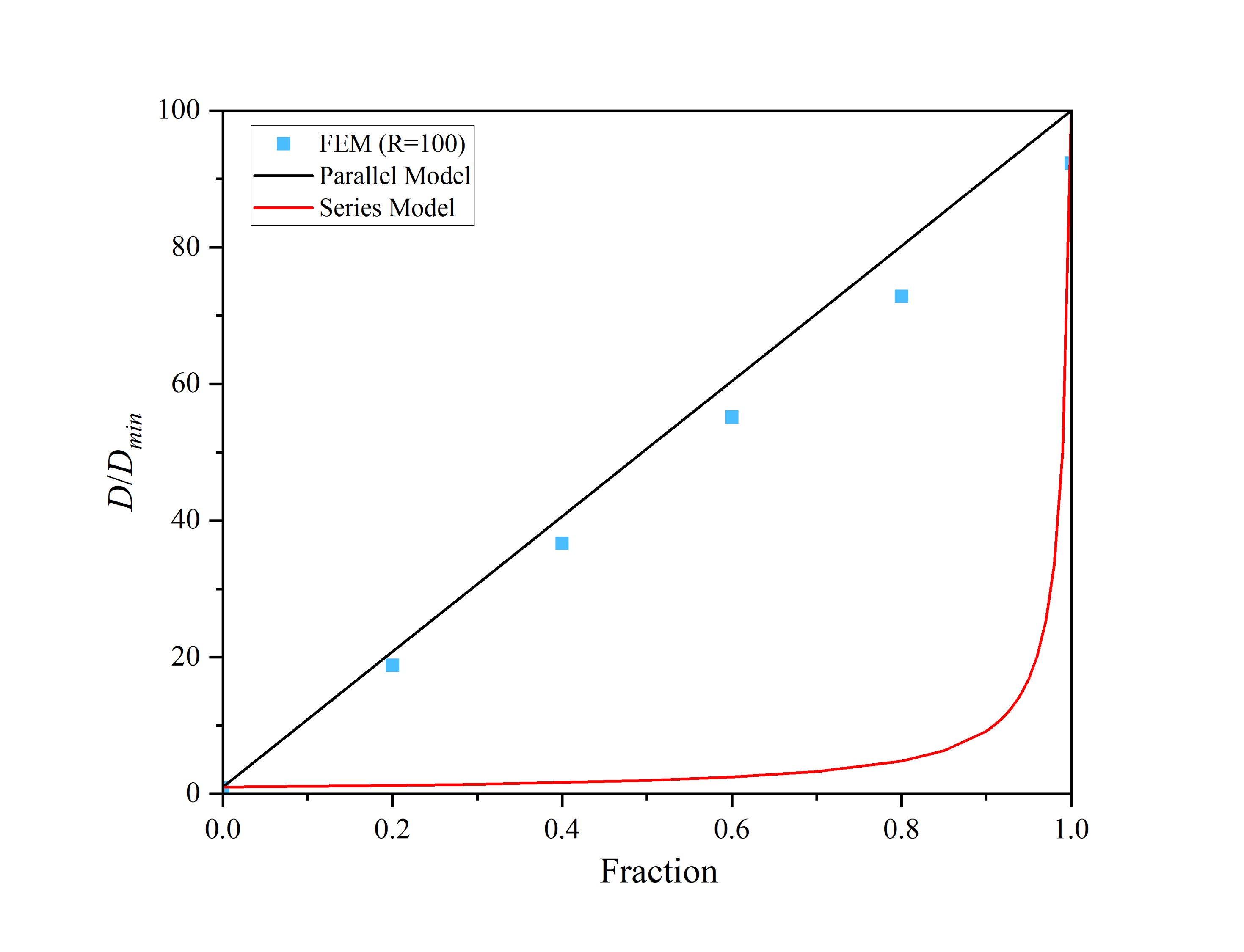}
  }
  \caption{ Calculated total diffusion coefficients of Model 1-2-1 in $x$ direction along with the parallel and series model (a) R=2 (b) R=10 (c) R=60 (D) R=100.}
  \label{fig:Details of FEM result and parallel and series model D11} 
\end{figure}

Fig~\ref{fig:Details of FEM result and parallel and series model D11} shows the comparison of calculated total diffusion coefficients and predicted model in the $x$ direction for R of $2$, $10$, $60$, and $100$.

\newpage
\renewcommand\thefigure{\thesection.\arabic{figure}}
\setcounter{figure}{0}    

\section{Simulation of Moisture Diffusion in the $z$ direction ($D_{z}$)}

\begin{figure}[h!] 
  \centering
  \subfigure[]
  {
      \label{fig:3D Map D33 model 2-1-2}\includegraphics[width=0.48\linewidth]{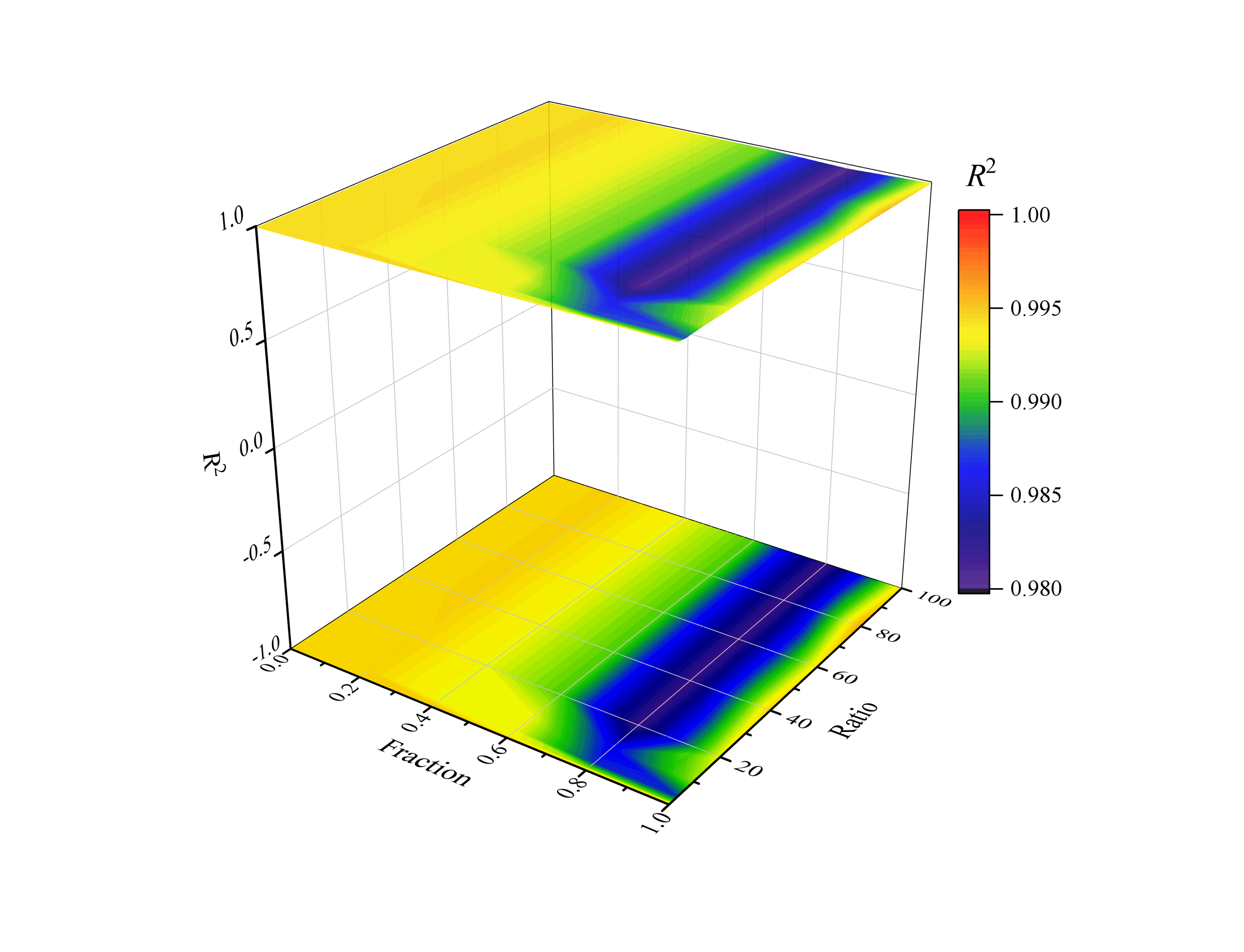}
  }
  \subfigure[]
  {
      \label{fig:Boundary map D33 model 2-1-2}\includegraphics[width=0.48\linewidth]{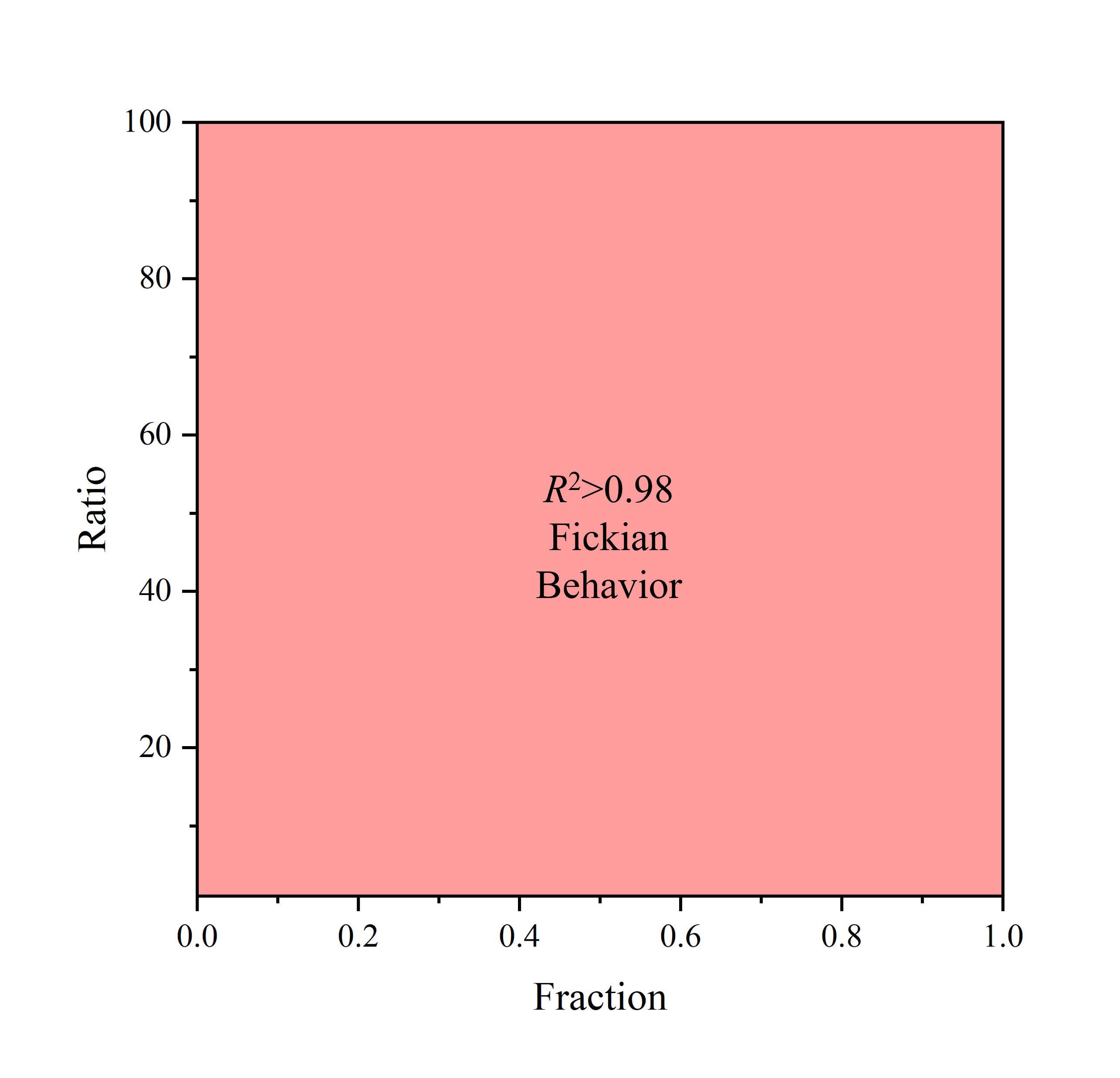}
  }
  \caption{ Results of linear regression analysis of the simulation data from Model 2-1-2 in the $z$ direction (a) Three dimension contour map of $R^2$ with respect to R and F (b) Boundary map of $R^2$ with respect to R and F indicating clear regions of Fickian behavior.}
  \label{fig:R2 model 2-1-2} 
\end{figure}

\begin{figure}[h!] 
  \centering
  \subfigure[]
  {
      \label{fig:D33 model 2-1-2 R=2}\includegraphics[width=0.48\linewidth]{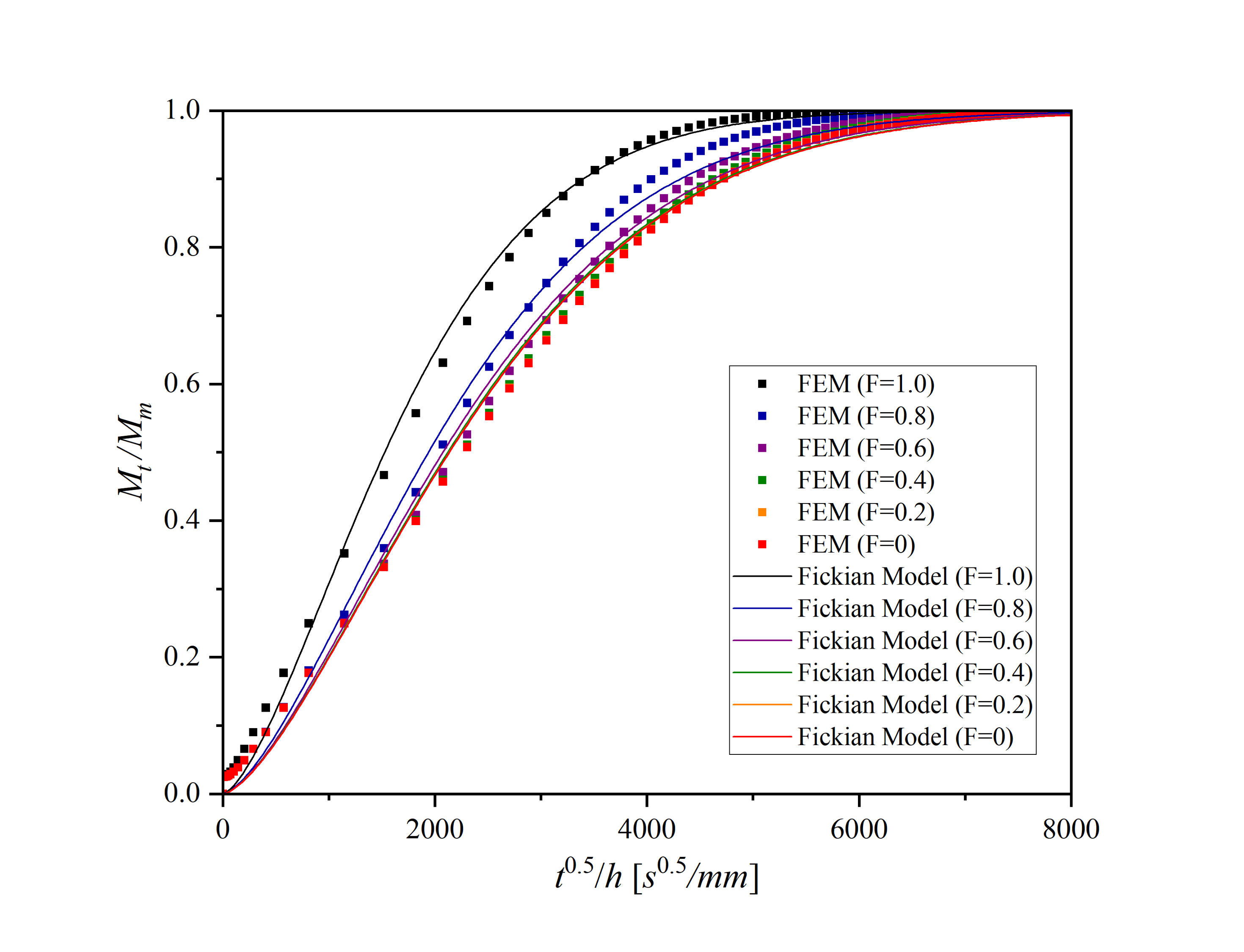}
  }
  \subfigure[]
  {
      \label{fig:D33 model 2-1-2 R=10}\includegraphics[width=0.48\linewidth]{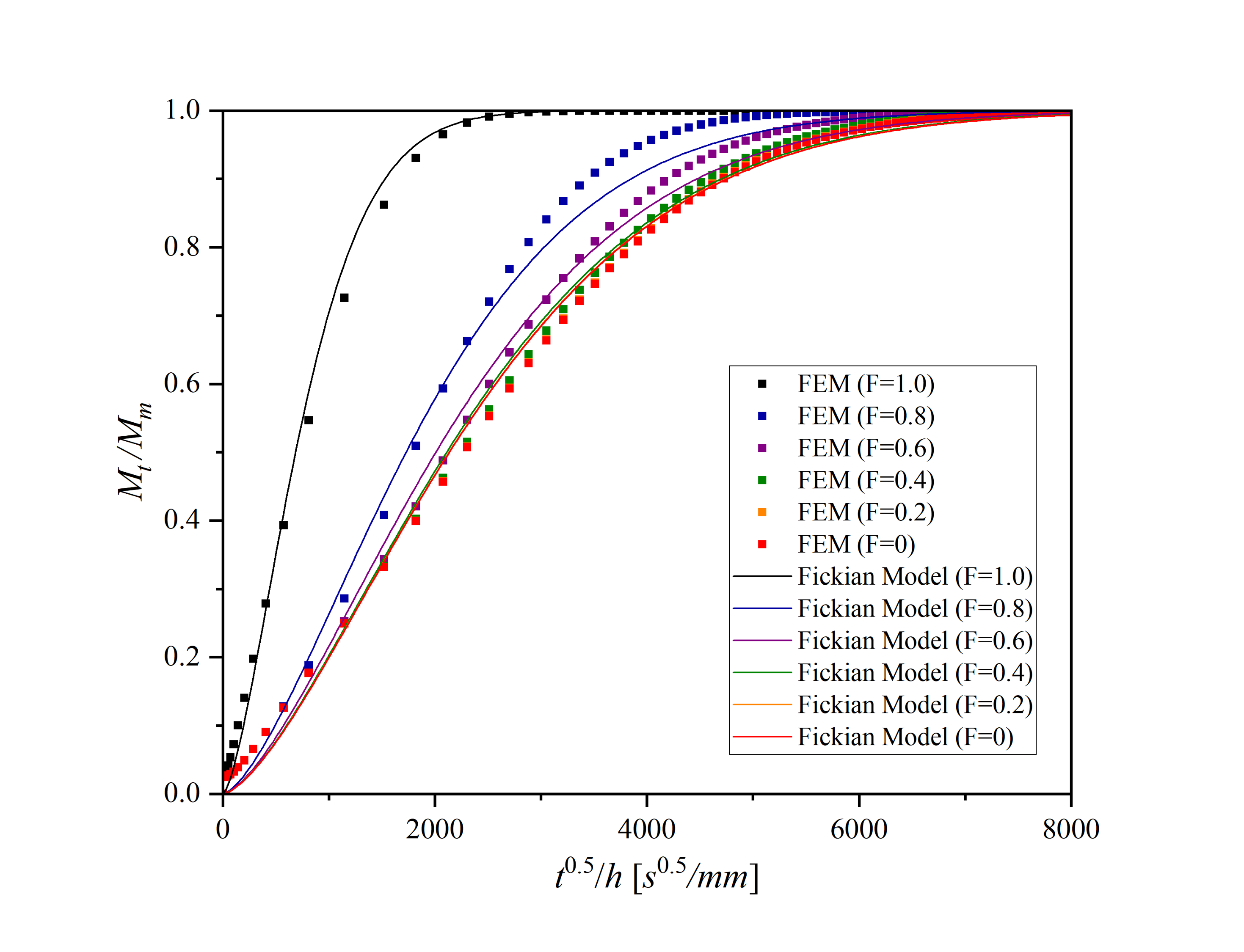}
  }
    \subfigure[]
  {
      \label{fig:D33 model 2-1-2 R=60}\includegraphics[width=0.48\linewidth]{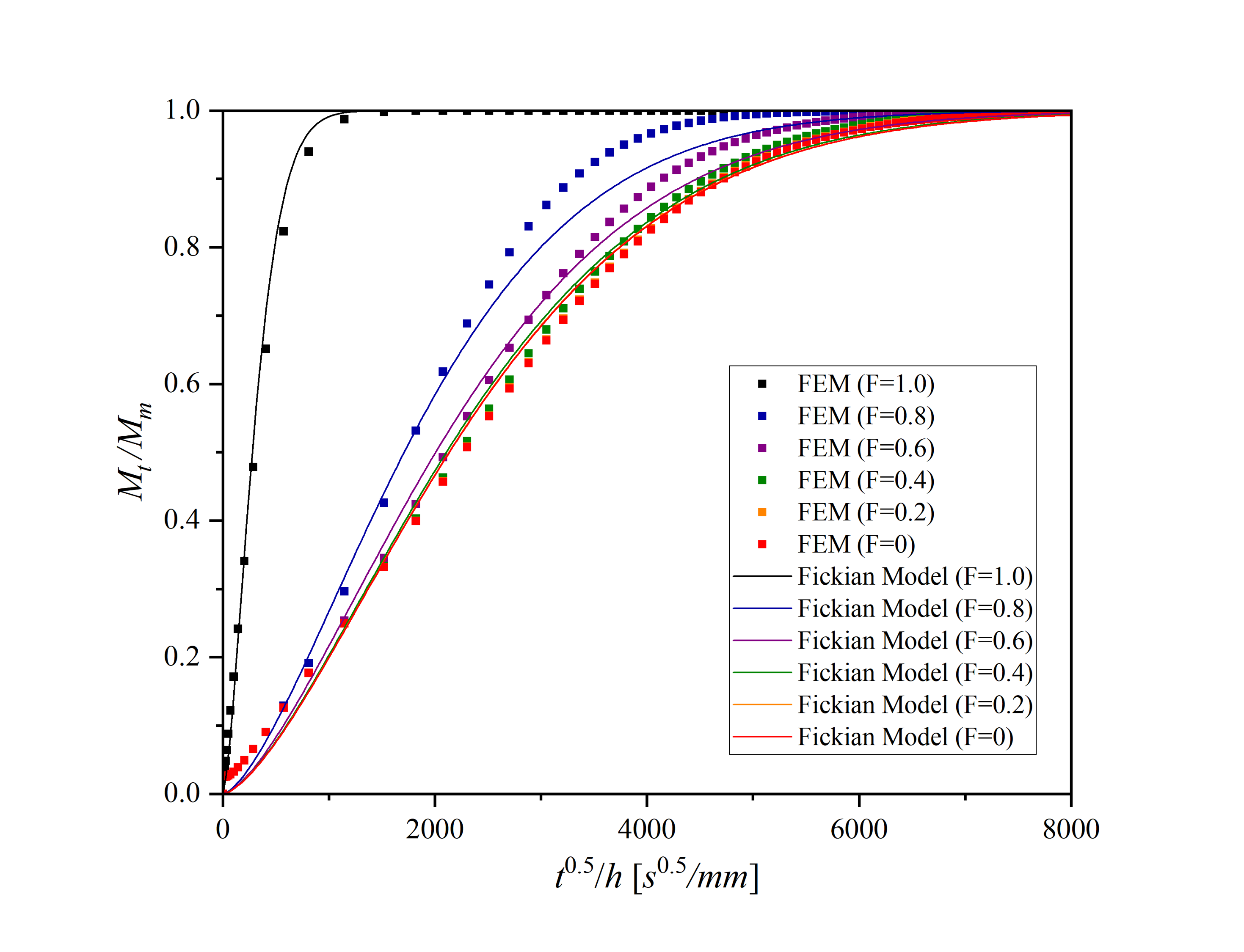}
  }
  \subfigure[]
  {
      \label{fig:D33 model 2-1-2 R=100}\includegraphics[width=0.48\linewidth]{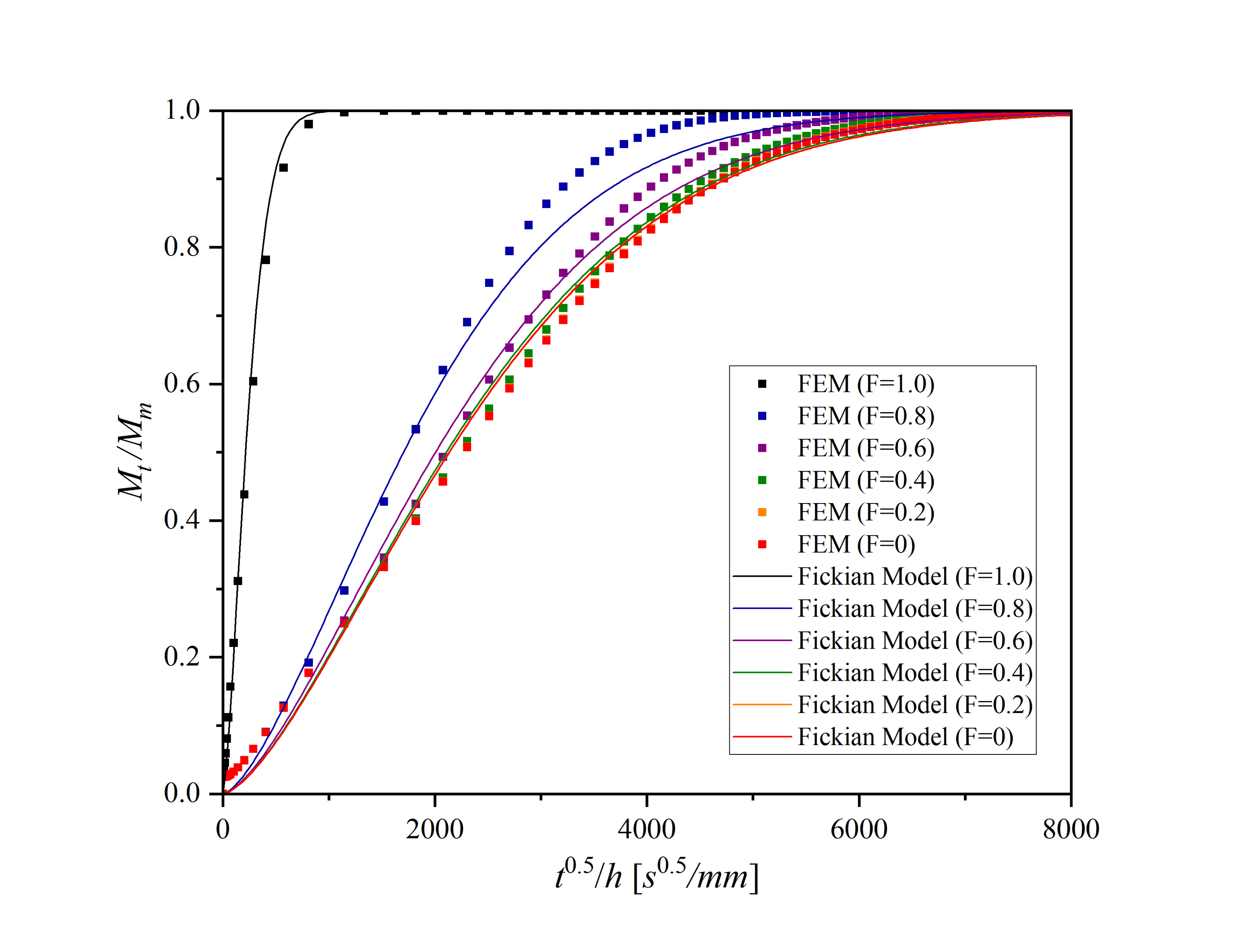}
  }
  \caption{ Results of simulation data and Fickian equation fit for Model 2-1-2 in the $z$ direction (a) R=2 (b) R=10 (c) R=60 (d) R=100.}
  \label{fig:FEM result D33 model 2-1-2} 
\end{figure}

Fig~\ref{fig:FEM result D33 model 2-1-2} presents the result of Model 2-1-2 for R of $2$, $10$, $60$, and $100$, wherein the scatter plots represent the FEM simulation data. Continuous lines delineate the predictions based on the analytical Fickian model given in Eq~\ref{eq:linear model} and Eq~\ref{eq:approximation model}.

\end{document}